%% file: sample-manuscript.tex
\documentclass[sigconf]{acmart}

\AtBeginDocument{%
  }

\setcopyright{acmlicensed}\acmConference[UIST '25]{The 38th Annual ACM Symposium on User Interface Software and Technology}{September 28-October 1, 2025}{Busan, Republic of Korea}
\copyrightyear{2025}
\acmYear{2025}
\acmDOI{10.1145/3746059.3747670}

\acmBooktitle{The 38th Annual ACM Symposium on User Interface Software and Technology (UIST '25), September 28-October 1, 2025, Busan, Republic of Korea}
\acmISBN{979-8-4007-2037-6/2025/09}

\newcommand{\systemName}{{CoGrader}}

\newcommand{\revision}[1]{\textcolor{black}{#1}}

\usepackage{cleveref}
\usepackage{multirow}

\acmSubmissionID{1641} 

\usepackage{colortbl}
\definecolor{lightBlue}{HTML}{A7C8DE}
\definecolor{DarkBlue}{HTML}{5790C0}
\definecolor{DarkGreen}{HTML}{3d6f28}
\definecolor{DarkRed}{HTML}{b02102}
\definecolor{HighlightBlue}{RGB}{220,230,241}
\definecolor{HighlightRed}{HTML}{B22222}

\usepackage{graphicx}

\usepackage{acmart-taps}

\begin{document}

\title{CoGrader: Transforming Instructors' Assessment of Project Reports through Collaborative LLM Integration}

\author{Zixin Chen}
\email{zchendf@connect.ust.hk}
\orcid{0000-0001-8507-4399}
\affiliation{%
  \institution{The Hong Kong University of Science and Technology}
  \city{Hong Kong}
  \country{China}
}

\author{Jiachen Wang}
\email{wangjiachen@zju.edu.cn}
\orcid{0000-0001-9630-9958}
\affiliation{%
  \institution{Zhejiang University}
  \city{Hangzhou}
  \state{Zhejiang}
  \country{China}
}

\author{Yumeng Li}
\email{liyumengcafa@gmail.com}
\orcid{0009-0001-3286-030X}
\affiliation{%
  \institution{The University of Hong Kong}
  \city{Hong Kong}
  \country{China}}

\author{Haobo Li}
\email{hliem@connect.ust.hk}
\orcid{0009-0003-8771-2835}
\affiliation{%
    \institution{The Hong Kong University of Science and Technology}
  \city{Hong Kong}
  \country{China}
}

\author{Chuhan Shi}
\email{chuhanshi@seu.edu.cn}
\orcid{0000-0002-3370-1626}
\affiliation{%
 \institution{Southeast University}
 \city{Nanjing}
 \state{Jiangsu}
 \country{China}}

\author{Rong Zhang}
\email{rzhangab@connect.ust.hk}
\orcid{0000-0002-4669-4816}
\affiliation{%
  \institution{The Hong Kong University of Science and Technology}
  \city{Hong Kong}
  \country{China}}

\author{Huamin Qu}
\email{huamin@cse.ust.hk}
\orcid{0000-0002-3344-9694}
\affiliation{%
  \institution{The Hong Kong University of Science and Technology}
  \city{Hong Kong}
  \country{China}}

\begin{abstract}
  Grading project reports is increasingly significant in today's educational landscape, where they serve as key assessments of students' comprehensive problem-solving abilities. However, it remains challenging due to the multifaceted evaluation criteria involved, such as creativity and peer-comparative achievement. Meanwhile, instructors often struggle to maintain fairness throughout the time-consuming grading process. Recent advances in AI, particularly large language models, have demonstrated potential for automating simpler grading tasks, such as assessing quizzes or basic writing quality. However, these tools often fall short when it comes to complex metrics, like design innovation and the practical application of knowledge, that require an instructor's educational insights and contextual understanding of the class. To address this challenge, we conducted a formative study with six instructors and developed {\systemName}, which introduces a novel grading workflow combining human-LLM collaborative metrics design, benchmarking, and AI-assisted feedback. {\systemName} was found effective in improving grading efficiency and consistency while providing reliable peer-comparative feedback to students. We also discuss design insights and ethical considerations for the development of human-AI collaborative grading systems.
\end{abstract}



\maketitle

\input{sections/01-introduction}

\input{sections/02-related-work}

\input{sections/03-formative-study}

\input{sections/04-design}
\input{sections/05-evaluation}

\input{sections/06-discussion}
\input{sections/07-conclusion}
\input{sections/09_ack}

\bibliographystyle{ACM-Reference-Format}
\bibliography{sample-base}

\end{document}

%% file: sections/01-introduction.tex
\section{Introduction}

With the revolutionary development of generative artificial intelligence (GAI), not only is our day-to-day life undergoing significant changes, but the skills that are valued and emphasized in education are also shifting in a foreseeable way~\cite{bahroun2023transforming,farrelly2023generative}. Recent educational research argues that skills such as critical thinking, creativity, and problem-solving are becoming increasingly important, as they enable individuals to move beyond simple information recall and engage in complex, valued thinking processes~\cite{liu2023future,tsai2023exploring,alasadi2023generative}. Among various learning activities, project-based learning (PBL) stands out as particularly effective in cultivating these significant thinking skills~\cite{maida2011project,bender2012project,albar2021problem}. By immersing students in real-world challenges, PBL fosters the application of theoretical knowledge, the development of original ideas, and the synthesis of interdisciplinary insights~\cite{chen2022effect,sasson2018fostering}. 


Despite its educational value, assessing project reports remains one of the most challenging and time-consuming tasks for instructors~\cite{timmis2016rethinking,alves2009teachers}. Unlike standardized learning activities such as quizzes, project reports require evaluation across a range of complex, multifaceted, and often customized metrics that rely on an instructor's educational insights, including assessments of originality, practical application of knowledge, and the level of effort and achievement compared to peers~\cite{menzies2016project,beckett2002teacher}. The complexity of these metrics, combined with the need for subjective judgment, necessitates nuanced evaluation and deep domain expertise, making the grading process particularly demanding. Additionally, instructors face the challenge of maintaining consistency and fairness throughout the grading process, especially over extended periods where there is a risk of grading standards drifting~\cite{walvoord2011effective,nilson2015specifications,rasooli2018re}. In many cases, instructors must revisit and adjust earlier scores as they develop a clearer sense of overall class performance, resulting in repetitive work and potential inefficiencies~\cite{wormeli2023fair}.

Recent advances in AI, particularly large language models (LLMs) such as ChatGPT, have demonstrated significant promise in automating various educational tasks for instructors~\cite{meyer2023chatgpt, tian2024opportunities, abd2023large}. For example, LLMs have successfully been employed to generate multiple-choice questions, provide personalized exercises and feedback to students, and even support the training of novice teachers~\cite{olney2023generating, pan2025tutorup, tran2023generating, meyer2024using, song2024automated}. Specifically in the context of grading, existing studies highlight LLMs' strengths in efficiently evaluating standardized components such as structural clarity, coherence, and basic language correctness~\cite{sidorkin2024embracing, kostic2024llms}. However, LLMs often struggle when grading complex or subjective criteria, such as assessing originality, evaluating creative problem-solving approaches, or accurately interpreting context-specific insights in project reports. Common issues include misinterpreting nuanced student arguments, providing overly generic feedback, or incorrectly assessing the significance of unique ideas or methods presented by students~\cite{bernabei2023students, li2023adapting}.

Conversely, experienced instructors excel in making nuanced judgments informed by their educational insights and contextual knowledge of the class, yet they can experience challenges in maintaining consistent standards or managing potential biases over extensive grading tasks. Thus, a collaborative approach where human instructors retain primary responsibility for assessment, supported by AI-generated suggestions and preliminary evaluations, provides an effective solution. Human oversight ensures accuracy, fairness, and customized feedback specifically aligned with the unique qualities and learning objectives of each student's project report~\cite{moore2023empowering, de2023chatgpt, atlas2023chatgpt}.

As a result, while LLMs has made significant strides in automating educational tasks, its integration into the complex workflow of grading project reports remains largely unexplored. To address this gap, we propose a human-AI collaborative workflow that blends LLMs' computational strengths with human instructors' nuanced expertise. The workflow begins with an LLM-powered project requirement analysis and collaborative metric design, allowing instructors to select and refine relevant report assessment metrics. Next, LLM automatically grades the reports, providing reference scores and comments for each user-selected metric. Instructors then review these initial assessments, using LLM-generated insights to guide the selection of benchmark reports. Once benchmarks are established, LLM re-evaluates each report, offering comments that compare the performance on each metric against these benchmarks. This not only provides students with individualized feedback but also allows them to learn from peer comparisons by understanding how their work measures up to the benchmark reports. Instructors can review and modify LLM’s comments for each criterion, while also freely adding their own annotations and comments on other parts of the report as needed. Finally, LLM consolidates all grades and comments, generating personalized, comprehensive feedback for each report.

Throughout the process, instructors retain full authority over judgments, while the LLM accelerates the grading workflow by providing initial evaluations and speeding up benchmark-driven regrading, all while ensuring detailed feedback for students. This collaborative approach combines AI efficiency with human expertise, maintaining fairness and accuracy with the instructors’ judgment at the core. As a result, students receive more timely, peer-comparison-based, and personalized feedback, enabling them to reflect on their work and learn from their peers, fostering growth and deeper understanding in project-based learning. To summarize, our paper makes the following three contributions:

\begin{itemize} 

\item A formative study that identifies instructors' challenges in grading project reports and explores how integrating LLMs like ChatGPT can address these challenges while keeping instructors in control, laying the foundation for an effective human-AI co-grading framework.

\item The design and implementation of {\systemName}, a novel human-AI collaborative grading workflow that integrates LLMs for metric design, report evaluation, benchmark comparison, and feedback generation, enhancing grading efficiency and consistency while providing students with valuable, peer-comparative feedback to support their learning and improvement.

\item An evaluation of {\systemName} demonstrates the effectiveness of the proposed workflow, highlights instructors' strong preference for its use, and underscores the broader potential of human-AI collaboration in tackling complex educational tasks, ultimately enhancing students' learning experiences.

\end{itemize}

%% file: sections/02-related-work.tex
\section{Related Work}

In this section, we review prior work on the challenges of assessing project-based learning (PBL), AI-powered tools for grading and educational assessment support, and the use of large language models (LLMs) as interfaces to assist instructors in educational contexts.

\subsection{PBL and Assessment Challenges}
PBL is widely regarded as an effective pedagogical method by engaging students in real-world~\cite{krajcik2006project} and interdisciplinary~\cite{brassler2017enhance} challenges. In PBL, students typically engage in long-term projects, acquiring practical skills through completing artifacts, such as project reports~\cite{phyllis1991motivating}. 
Research has explored the advantages of PBL in enhancing students' skills, including fostering self-regulated learning ~\cite{dimitra2016project}, cultivating creativity~\cite{guo2020review}, and developing critical thinking and collaboration~\cite{masek2011effect}.

Grading project reports is one of the primary method for assessing the learning outcomes of PBL. The traditional grading method utilizes a comprehensive grading rubric that outlines specific criteria and corresponding levels of achievement for each aspect of the project~\cite{brodie2009comparison, hashemi2024llm}. While this method aims to standardize evaluations, it has limitations.
First, the rubric makes it hard to capture qualitative elements like creativity and originality~\cite{clary2011measuring}.
Second, rubrics can introduce bias~\cite{panadero2020critical} in evaluation when multiple instructors or teaching assistants review reports, as the evaluators' personal preferences can influence the grading process.
Third, developing a standard and comprehensive rubric for PBL projects by instructors can be time-consuming, especially when dealing with complex projects or diverse student work~\cite{panadero2020critical}. Moreover, rubric-based grading is susceptible to grading drift over time~\cite{leckie2011rater}, as instructors often find it difficult to maintain consistent and fair evaluations throughout a lengthy grading process.

These limitations pose significant challenges due to the need for instructors to evaluate diverse, complex, and often subjective rubrics and metrics. Our proposed workflow addresses these challenges through a human-AI collaborative approach. By combining collaborative metric design and selection with an iterative, LLM-facilitated grading process, instructors can reduce the burden of grading standardized, repetitive metrics, allowing greater focus on complex, subjective evaluations. Meanwhile, the benchmarking-driven auto regrading process significantly accelerates the grading workflow, promotes fairness and mitigates grading drift, ensuring consistent and reliable assessments.



\subsection{AI for Learning Assessment}
The increasing complexity of educational assessment has driven the development of AI tools to support grading and feedback. These tools typically either automate the grading process or assist instructors by providing references and insights in a human-AI collaborative manner.


Advancements in artificial intelligence have enabled the development of tools that automate grading and assessment tasks~\cite{hooper2024advancing}. For example, AI-powered programming assistants such as CodeAid and Learning with Style~\cite{kazemitabaar2024codeaid,saliba2024learning} offer automated, real-time evaluation of code quality and style. Similarly, tools like VoiceCoach and TypeTalker~\cite{wang2020voicecoach,arawjo2017typetalker} provide immediate grading and comments for students’ spoken language assignments. While these tools significantly enhance efficiency and consistency in assessment, they also raise concerns around transparency, fairness, and potential biases in AI-driven evaluation~\cite{mcconvey2023human, zheng2023competent}, underscoring the importance of maintaining human oversight in educational contexts.


In addition to automated grading tools, a growing body of work explores the use of AI to facilitate feedback generation and communication among students and instructors, recognizing the pedagogical value of peer evaluation and teacher feedback~\cite{yuan2016almost, schank2004collaborative}. Tools such as SlideSpecs~\cite{warner2023slidespecs} and chat.codes~\cite{oney2018creating} help aggregate presentation feedback and code comments from student discussions, streamlining the feedback process and enhancing communication efficiency.

Other platforms—such as RichReview++~\cite{yoon2016richreview++}, edX~\cite{yoon2015multi}, GroupMeter~\cite{leshed2009visualizing}, and DeepThinkingMap~\cite{liao2023deepthinkingmap}—employ multimodal AI techniques to support feedback generation through visualizations, annotations, and recommendation systems for both peers and instructors. Additionally, large-scale peer grading and peer-comparison tools have been developed to enhance fairness and scalability in assessing subjective assignments~\cite{coetzee2015structuring, zhang2024peer, park2017eliph, glassman2016learnersourcing}. While these systems highlight the benefits of feedback in fostering learning and engagement, they often rely heavily on active student participation, which can limit their effectiveness and scalability in real-world classroom settings.

\revision{Further work highlights the trade-offs in feedback quality and timing, especially in balancing rapid feedback with depth and personalization. Studies have explored how feedback timing impacts learning outcomes~\cite{ngoon2018interactive,yen2024give} and how students perceive automatically generated feedback~\cite{bayerlein2014students}. These insights reinforce the importance of designing AI-assisted feedback systems that maintain pedagogical value while providing timely support.}

Our work builds on these foundations by introducing an AI-assisted framework that automates large-scale, peer-comparative evaluations under human guidance. The system ensures that instructors retain a central, decision-making role in the grading workflow through clearly defined responsibilities, while leveraging AI to enhance efficiency and consistency. It also promotes transparency and generates concise, personalized feedback for students. This approach addresses key ethical concerns in AI-driven assessment by preserving instructor oversight and reinforcing the educational value of meaningful, actionable feedback.

\subsection{LLM-Powered Educational Support Systems}


Recent studies have increasingly leveraged large language models (LLMs) to build educational systems that support both student learning and instructor workflows~\cite{chen2024learning,ross2023programmer,daryanto2025conversate,jin2024teach,shao2025unlocking}. In this section, we focus on the latter—LLM-powered tools designed to assist instructors with a range of other educational tasks beyond grading.

For example, LLMs have been used to assist with course planning~\cite{fan2024lessonplanner,hu2024teaching}, project or assignment design~\cite{ravi2025co}, and evaluating students' usage of generative AI tools~\cite{zheng2024selfgauge}. Other systems leverage LLMs for interdisciplinary ideation~\cite{fan2025litlinker} or simulate virtual classmates and teaching assistants to facilitate classroom interactions and collaborative learning~\cite{liu2024classmeta,kazemitabaar2024codeaid,kumar2024guiding}.


Additionally, several systems have integrated LLMs into learning analytics pipelines and formative feedback workflows, enabling personalized support based on students’ behavior and performance~\cite{tang2024vizgroup,jia2024llm,stamper2024enhancing,zhang2025cpvis,chen2024stugptviz}. These works highlight the growing role of LLM-powered interfaces in enhancing various aspects of instructional design, classroom facilitation, and teaching support.


\revision{However, recent studies have raised concerns that LLM-powered systems may inadvertently reduce instructor engagement and cognitive effort over time~\cite{lee2025impact,kumar2025human}. Additionally, emerging research has shown that LLMs may exhibit ``self-favoritism'' biases, rating their own outputs more favorably than human-written content~\cite{panickssery2024llm}, posing risks to fairness in grading contexts where originality is critical. Our system mitigates these concerns by positioning AI-generated comments primarily as evidential support rather than as final judgment. Key procedures, such as benchmark selection and AI comment revision, remain fully instructor-driven to encourage critical review of AI outputs and ensure fairness through sustained human oversight.}

Building on these efforts, our work addresses a more complex and pedagogically significant challenge: grading open-ended reports in project-based learning. It advances human-AI collaboration by providing a scalable solution that supports instructors in managing grading and helps students receive meaningful, improvement-oriented feedback.

%% file: sections/03-formative-study.tex
\section{Formative Study}
\label{sec:formative-study}

To gain a deeper understanding of the challenges and needs instructors face when grading students' project reports, we conducted semi-structured interviews with six university instructors, including two assistant professors, three lecturers, and one postdoctoral researcher. All participants had at least three years of experience in designing and grading project reports. 


\revision{Each session lasted approximately 70 minutes and included two parts: a 30-minute discussion focused on current project report grading workflows and their pain points, followed by a 40-minute brainstorming session to identify which evaluation aspects might be appropriate for AI automation and which should remain under human control. This structure helped surface both practical challenges and design opportunities for collaborative AI-assisted grading systems.}

\revision{With participant consent, three co-authors independently reviewed the interview notes and recordings. Using open coding, they extracted key concepts and clustered them into emerging themes. Through two rounds of discussion, discrepancies were resolved, and consensus was reached on the core themes. Below, we summarize the key findings from these discussions.}



\subsection{Workflow and Challenges in Grading Project Reports} Instructors generally follow a structured workflow when grading project reports, though they encounter several recurring challenges at each stage. Insights from our interviews reveal the following workflow stages and associated challenges:


    \textbf{S1: Appropriate Grading Metrics Design:} The grading process begins with designing metrics that align closely with course learning outcomes and project requirements. A primary challenge at this stage is ensuring that these metrics remain consistently aligned with the objectives shared with students, providing clear expectations while accommodating diverse approaches in student submissions.

    \textbf{S2: Quick Overview and Benchmark Selection:} Instructors often start the grading process by skimming through all submissions to gain a high-level understanding of the overall quality and performance distribution within the class. During this phase, they also identify a few representative submissions to serve as benchmarks to guide more consistent evaluation later. However, this step is always cognitively demanding, especially when the number of submissions is large or when the quality differences are subtle. Instructors often find it challenging to quickly identify suitable benchmark reports while ensuring their initial impressions remain unbiased and representative of the overall submission quality.

    \textbf{S3: Detailed Evaluation, Benchmark Adjustments, and Retroactive Grading:} After the initial overview and selection of benchmark reports, instructors proceed with the detailed evaluation of each report according to the grading metrics, often comparing them to the selected benchmarks. During this phase, instructors frequently update benchmarks and revisit previously graded reports to ensure consistency, making necessary adjustments based on new insights. This iterative process of retroactive grading is time-consuming and poses significant challenges, particularly when grading open-ended reports. Instructors highlighted the challenge of maintaining fairness in subjective judgment throughout the grading process, especially when evaluating factors like innovation and peer-comparative design quality.

    \textbf{S4: Comprehensive Feedback and Synthesizing Class-Level Insights:} After evaluating each report, instructors provide feedback, though they often find it challenging to deliver detailed, cohesive comments that address all critical points. Writing comprehensive feedback is time-intensive, leading some instructors to focus on limited key points and encourage students to seek additional guidance in office hours. Instructors also recognize the importance of comparing individual reports against overall class performance to highlight improvement areas. However, synthesizing these insights across all reports and effectively communicating them to the class presents another significant challenge, requiring substantial time and effort to ensure feedback is both meaningful and actionable.

\subsection{Perceptions of AI in Grading Assistance}


In the second phase of our formative study, we first asked instructors to share their overall impressions of integrating AI tools like ChatGPT into the grading process. All instructors expressed an open-minded and positive attitude toward using AI to support the evaluation of project reports. At the same time, most emphasized the necessity of maintaining a certain level of human control over key parts of the grading workflow. To gain deeper insights into the appropriate roles of AI in the system, we further posed two guiding questions and encouraged instructors to openly brainstorming:

\textit{(Q1) Which parts of the grading process could be effectively managed by AI tools alone such as ChatGPT or automated algorithms?}

\textit{(Q2) Which aspects of grading would remain challenging or unreliable for AI alone, where human expertise is indispensable?}

Based on their reflections, we identified specific tasks corresponding to each question. In the following section, we first summarize the tasks that instructors viewed as well-suited for AI automation (Q1), followed by tasks where AI could play a supporting role but human judgment remains essential (Q2).


\subsubsection{Tasks Suitable for AI and Automated Algorithms Alone:}

\begin{enumerate} 
\item \textbf{Automated Metrics Generation and Recommendation:} AI can support instructors by analyzing project requirements to suggest relevant assessment metrics, acting as a brainstorming tool to recommend valuable criteria for consideration in grading design.
\item \textbf{Standardized Report Evaluation:} Instructors noted AI's potential to handle basic assessments of submissions, including generating summary statistics, preliminary scores, and comments based on standardized criteria. These automated checks could cover areas like formatting, grammar, and other objective elements.
\item \textbf{Content Summarization and Key Point Extraction:} AI can assist by efficiently summarizing reports and extracting key points, providing instructors with a quick and clear overview of each submission’s core ideas.
\end{enumerate}


\subsubsection{Tasks Where AI Assists but Human Judgment is Essential:}

    \begin{figure*}[!tbp]
  \includegraphics[width=\textwidth]{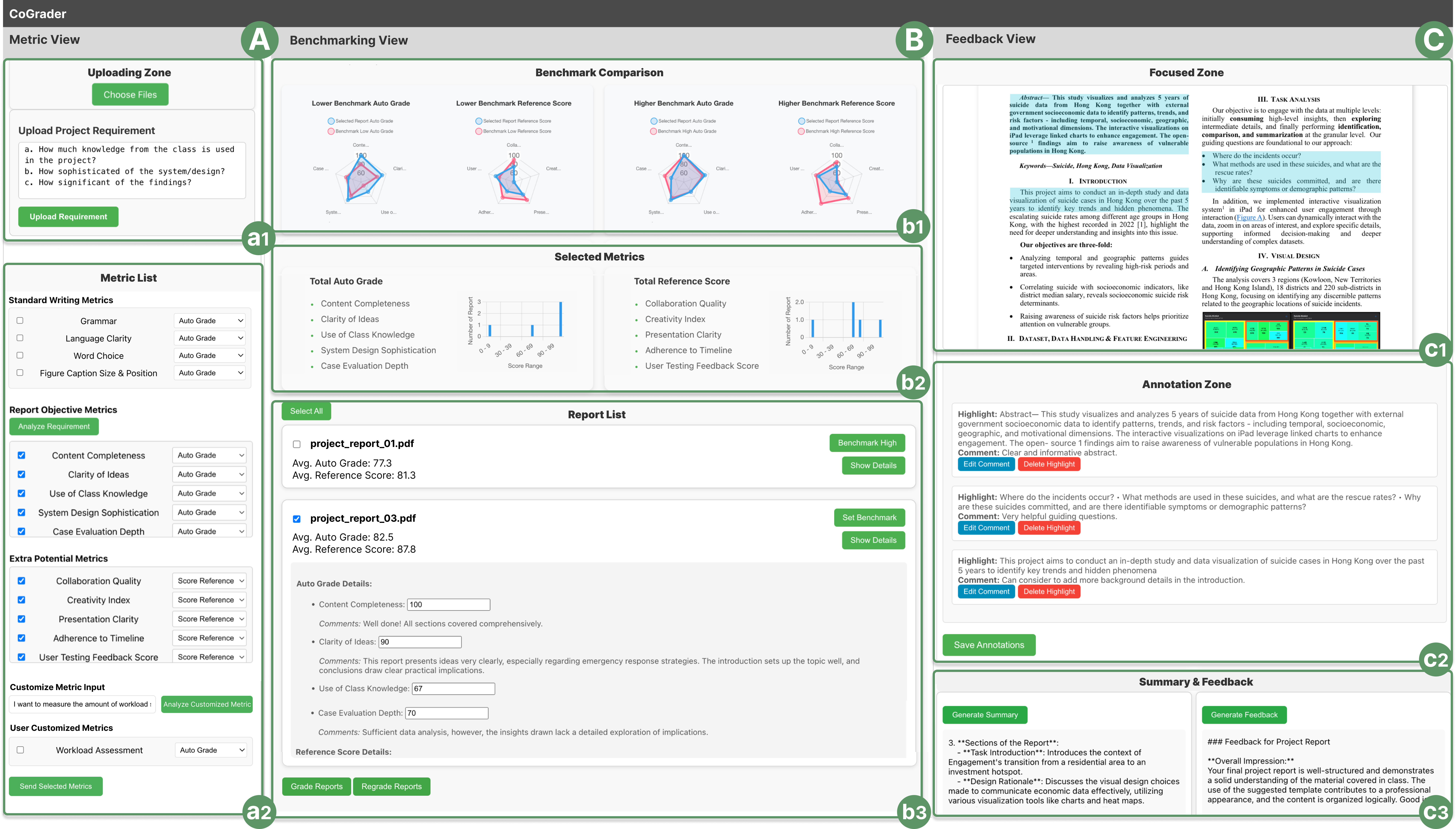}
  \caption[]{\revision{System interface of {\systemName}.} The \textit{Metric View} (A) provides functions for uploading reports and project requirements (a1), along with co-design tools for generating multi-perspective project assessment metrics (a2). The \textit{Benchmarking View} (B) includes a Benchmark Comparison zone (b1) where users can compare metric scores between a focused report and user-selected benchmarks. The Selected Metrics zone (b2) below offers an overview of metric score distribution across all reports. The Report List zone (b3) displays uploaded reports, allowing users to trigger automatic report evaluation, review and edit AI-generated scores and comments for each metric, or select benchmarks for AI-assisted iterative regrading. The \textit{Feedback View} (C) shows the contents of the focused report (c1), enabling users to review and annotate (c2). Users can also access the AI-generated summary and create the final feedback for the focused report (c3).
  }
  \Description{The teaser figure.}
  \label{fig:teaser}
\end{figure*}

\begin{enumerate}

    \item \textbf{Benchmark Selection and Retroactive Grading:} 
    

    Instructors noted that while AI can streamline the benchmarking process by suggesting rankings, classifications, and comparisons among student reports, it may lack the nuanced understanding needed for complex assessments. AI can assist by automating score updates during retroactive grading, reducing the manual effort required for revisiting prior evaluations. However, instructors emphasized that human judgment is essential to interpret context-specific elements and ensure that benchmarks align with course objectives and standards, particularly when evaluating complex or open-ended submissions.


    \item \textbf{Evaluating Subjective Aspects:} While AI can aid in evaluating subjective aspects—such as creativity, novelty, and aesthetic quality—final judgment should remain with instructors. AI could provide supportive comments and, where possible, generate quantitative insights to complement instructor evaluations, such as assessing workload intensity, identifying unique insights, or highlighting innovative ideas within reports.
    \item \textbf{Feedback and Insights Generation:} AI can assist in generating detailed, individualized feedback by organizing and expanding upon instructors' notes and evaluations, rather than generating feedback independently. This approach enables AI to provide students with clear, actionable points and compile class-wide insights to identify common strengths and areas for improvement. However, because AI lacks the contextual understanding and sensitivity required for nuanced feedback, human input remains essential to ensure the feedback aligns with learning objectives and addresses each student’s unique approach and progress.

\end{enumerate}

%% file: sections/04-design.tex
\section{System Design}
In this section, we first outline the core design requirements of the system, followed by an overview of {\systemName} through a representative user example. We then describe the system's frontend interface and backend components in detail.
\subsection{Design Requirements for the Grading Tool}
Based on the insights gained from the formative study, we identified several key requirements for the development of an AI-assisted grading tool that facilitates a human-AI collaborative process in grading students' project reports:

    \textbf{R1: Automatic Extraction of Core Metrics from Project Requirements.} The system should be able to analyze and interpret instructors’ project descriptions to automatically extract and summarize key grading criteria. This includes outlining essential evaluation metrics directly derived from the provided requirements, ensuring that foundational assessment standards are identified early. Automating this step helps streamline the initial grading setup and reduces manual workload, allowing instructors to concentrate on more complex and subjective aspects of evaluation.

    \textbf{R2: Enriching Metrics through Collaborative, Multi-Source Generation.} Building on the core metrics, the system should support a human-AI collaborative process to expand and refine the metric set. This includes AI-suggested metrics beyond project requirements, integration of external standards or rubrics, and instructor-provided additions or modifications. The system should also support co-designing new metrics tailored to specific learning goals, ensuring that the final assessment criteria are comprehensive, context-aware, and aligned with instructional intent.
    

    \textbf{R3: Automated Preliminary Analysis and AI-Driven Insights.} The system should provide automated preliminary assessments of all submissions, including summary statistics of evaluation scores and AI-suggested comments or recommendations for instructor review. These insights help instructors quickly grasp overall class performance, identify potential benchmark submissions, and enter the grading process with a clearer understanding of report quality distribution—ultimately supporting a more informed and efficient evaluation workflow.

    \textbf{R4: AI-Assisted Report Benchmarking, Comparison, and Retroactive Grading.} The system should support AI-assisted benchmarking and report comparison, allowing instructors to select, review, and update benchmark submissions with AI recommendations throughout the grading process. These benchmarks serve as reference points to maintain grading consistency. The tool should also enable automated retroactive grading, where previously scored reports are re-evaluated based on updated benchmarks. This AI-instructor collaboration reduces manual effort while ensuring fairness and coherence across all submissions.

    \revision{\textbf{R5: AI-Assisted, Instructor-Centered Feedback and Insight Generation.} The system should support instructors in composing clear, structured feedback by synthesizing their annotations, comments, and grading decisions across the evaluation process. AI-assisted summarization and phrasing can reduce the time and effort required to generate detailed feedback. Crucially, the final feedback should prioritize instructor-authored content, with AI-generated suggestions used to supplement areas not covered—ensuring clarity, consistency, and pedagogical integrity.}


    \revision{\textbf{R6: Ensure Instructor Authority and Balance AI Reliance.} The system must maintain instructors’ full authority over the grading process by clearly positioning AI outputs as supportive references rather than final decisions. AI should primarily assist with routine, low-judgment tasks—such as grammar checks and initial comment drafting—to reduce cognitive load and allow instructors to focus on nuanced, expertise-driven evaluation. All critical actions, including selecting criteria, identifying benchmark reports, and assigning final grades, remain under instructor control. To encourage instructor active review, AI-generated comments must include direct excerpts from student reports as evidence, discouraging passive acceptance of AI suggestions.}

Together, these design requirements respond to the key challenges surfaced in our formative study, highlighting opportunities where AI can meaningfully support and streamline the grading process, while ensuring that instructors retain full control over final decisions.


\subsection{{\systemName} System Design }
Building on the design requirements outlined above, we developed {\systemName}, a system that provides AI-driven support for instructors in grading project-based reports. {\systemName} is designed to seamlessly integrate into existing grading workflows, combining the efficiency of AI with the nuanced judgment of human instructors to enhance grading consistency and reduce workload. The system automates the analysis of project requirements, supports collaborative metric generation, and streamlines benchmarking and feedback processes. This section begins with a walkthrough of a typical user experience with {\systemName}, followed by a detailed explanation of each system component and how it addresses the identified design requirements.
\begin{figure*}[!tbp]
    \centering
    \includegraphics[width=\linewidth]{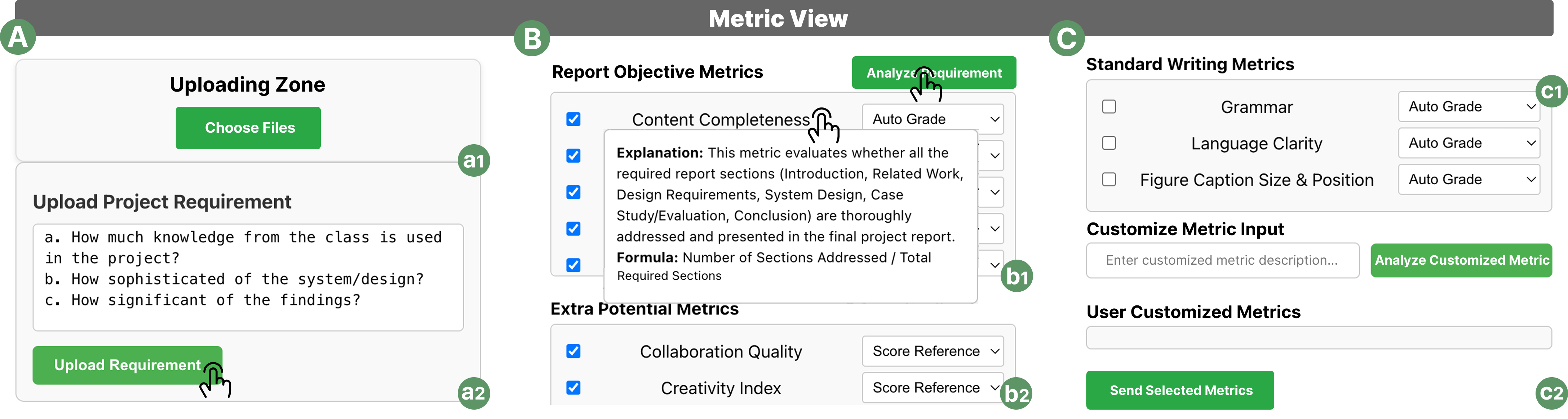}
    \caption[]{The components in \textit{Metric View}: (A) The Uploading Zone of {\systemName}, where users can upload project requirements and report files. (B) Automatic metric analysis, generating Report Objective Metrics and Extra Potential Metrics based on the uploaded project requirements. (C) The predefined Standard Writing Metrics and User Customized Metric design area which allow users to create metrics using natural language descriptions.}
    \label{fig:metricView}
\end{figure*}

\subsubsection{Example User Walkthrough.}
Jason teaches a data visualization course at a local university. As a hands-on course, he assigns a term project midway through the semester, which he plans to use as the primary assessment of his students' learning outcomes. The project is open-ended, allowing students to either choose from a set of sample topics or select their own datasets and perform self-directed analytics tasks. Along with the project, Jason provides students with a set of project requirements, which include key details such as the report style (e.g., academic writing), report contents (e.g., Introduction, Related Works, ...), and general grading metrics (e.g., system sophistication, design quality, ...).

At the end of the semester, Jason collects over 20 project reports, each approximately 5 pages long, and faces the daunting task of grading them. Although he has the project requirements, they are broad and abstract, making it challenging to grade efficiently while maintaining high-quality grading standards. To assist with this, Jason uses {\systemName}. Jason begins by uploading all the student reports into {\systemName}, along with the original project requirements that were shared with the students at the start of the term. He then clicks the ``Analyze Requirement'' button in the Metric List area, allowing the AI to assist in generating assessment metrics based on both the project requirements and AI suggestions. After reviewing the AI's analysis, Jason selects the recommended metrics: ``Content Coverage'', ``Technical Depth'', and ``Innovation and Creativity Index''. Recognizing that ``Content Coverage'' is more objective, he sets it as an ``Auto Grade'' metric. In contrast, he designates the other two as ``Score Reference'' metrics to reflect their more subjective nature and to remind himself to later review the AI-generated comments for those aspects.


With the metrics defined, Jason moves to the \textbf{Benchmarking View}, where he allows the AI to automatically grade all the reports based on the selected metrics and generate reference scores and comments. This provides him with an overview of the AI-recommended score distribution and the corresponding report rankings in the Report List below. By clicking on a report card, Jason can review its details in the right-hand \textbf{Feedback View}, where the Focused Zone allows for in-depth examination and reflection. He reviews the AI-generated scores and comments alongside the original report content, making annotations in the Annotation Zone and adjusting the scores and comments for each metric as needed. After reviewing about one-third of the reports, Jason identifies one report as a good example (higher benchmark) and another as a bad example (lower benchmark) . He designates these as benchmarks back in the Report List, spending additional time to refine the grading and comments for these two reports. To expedite the grading of the remaining reports, Jason selects them all and clicks the ``Regrade Reports'' button at the bottom of the Report List area, allowing the AI to regrade each report based on the previously defined metrics and the selected benchmarks. With the updated scores in hand, Jason conducts an iterative review, comparing each report's metric scores against the two benchmarks using the radar charts in the above Benchmark Comparison area. Once satisfied with the scores, Jason moves to the Summary \& Feedback zone in the bottom-right corner of the \textbf{Feedback View}, where he uses the AI to automatically generate personalized feedback based on the grading results, his annotations, and comments. Finally, Jason exports all the feedback and sends it to his students for their review.

\subsection{{\systemName} Frontend Components}
In the following section, we introduce the {\systemName} frontend user interface and the core components. 
\subsubsection{Metric View}
{\systemName} is meticulously designed to assist instructors in the project report grading process while ensuring they retain control throughout. The grading process begins with the instructor uploading the project report files (\cref{fig:metricView}-a1) and the project requirements (\cref{fig:metricView}-a2) via the Uploading Zone (\cref{fig:metricView}-A).

\begin{figure*}[!tbp]
    \centering
    \includegraphics[width=\linewidth]{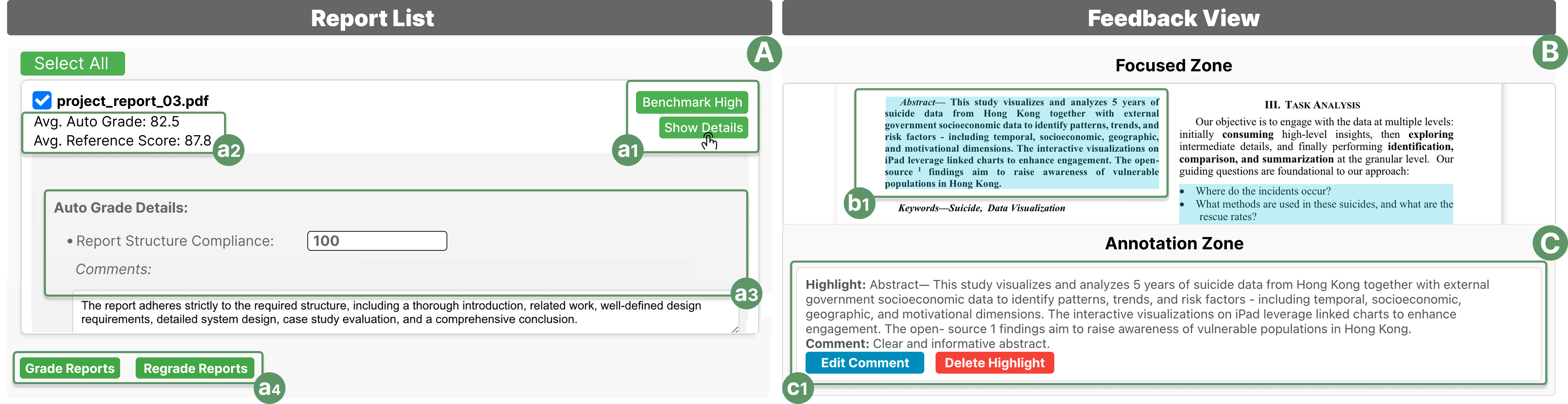}
\caption[]{(A) The Report List in the \textit{Benchmarking View}. Each uploaded report is presented as a report card. Users can view average scores for each metric group (a2), review and edit detailed scores and comments for selected metrics (a3), and select benchmark reports (a1). Additionally, users can initiate automatic grading and benchmark-driven regrading with a click (a4). (B \& C): The focused report details in the \textit{Feedback View}. (B) Users can view the focused report file by clicking on each report card in the report list (A). Users can freely highlight sections (b1) and add comments (c1) to the highlighted content. (C) These annotations are saved for use in generating report feedback.}
    \label{fig:reportList}
\end{figure*}

After the project requirements are uploaded, {\systemName} automatically analyzes these requirements to support the instructor’s grading metrics generation (\textbf{R1}). Instructors can initiate this process by clicking the ``Analyze Requirement'' button (\cref{fig:metricView}-b1). The system generates two categories of metrics: \textbf{Report Objective Metrics} (\cref{fig:metricView}-b1) and \textbf{Extra Potential Metrics} (\cref{fig:metricView}-b2). The \textbf{Report Objective Metrics} are derived directly from the uploaded requirements, ensuring alignment with the instructor's expectations. In contrast, the \textbf{Extra Potential Metrics} are suggested by the AI based on its understanding of valuable aspects that might not be explicitly mentioned in the requirements but could be important.



Each metric is presented in a list format, where hovering over the metric name provides a detailed explanation and suggested formula (\cref{fig:metricView}-b1). Instructors can select metrics for further analysis by checking the box next to each metric. They can also categorize each metric into two types: \textit{Auto Grade} or \textit{Score Reference}. \textit{Auto Grade} metrics are those where instructors can rely fully on AI-generated results, while \textit{Score Reference} metrics are more subjective, requiring AI suggestions but leaving the final judgment to the instructor. This design is in line with the user control principle outlined in \textbf{R6}, ensuring that from the metric selection phase onward, instructors have primary control while AI provides sufficient support.

Additionally, {\systemName} offers predefined \textbf{Standard Writing Metrics}, such as grammar and language clarity checks (\cref{fig:metricView}-c1), for instructors to select as needed. To provide even greater flexibility, the system includes a \textbf{Customized Metric Design} feature (\cref{fig:metricView}-c2), allowing instructors to describe additional metrics in natural language. The AI then assists by generating customized metrics and their corresponding formulas. It also cross-references the instructor's customized metrics with existing ones to avoid redundancy, highlighting any overlaps in red. This collaborative approach to metric design ensures that the generated metrics are comprehensive and reliable, aligning with the design requirement \textbf{R2} while still preserving user control as emphasized in \textbf{R6}.

\subsubsection{Benchmarking View \& Feedback View}
After co-designing and selecting the desired metrics, users can send the metrics to the \textbf{Benchmarking View} (\cref{fig:teaser}-B) for further report grading through a collaborative benchmarking process (\textbf{R4}). The \textbf{Benchmark Comparison} component (\cref{fig:teaser}-b1), the \textbf{Selected Metrics} component (\cref{fig:teaser}-b2) work together with the \textbf{Focused \& Annotation \& Summary Zone} (\cref{fig:teaser}-c1,c2,c3) in the \textbf{Feedback View} to assist instructors in benchmarking and reports grading.



\begin{figure}[!bp]
    \centering
    \includegraphics[width=\linewidth]{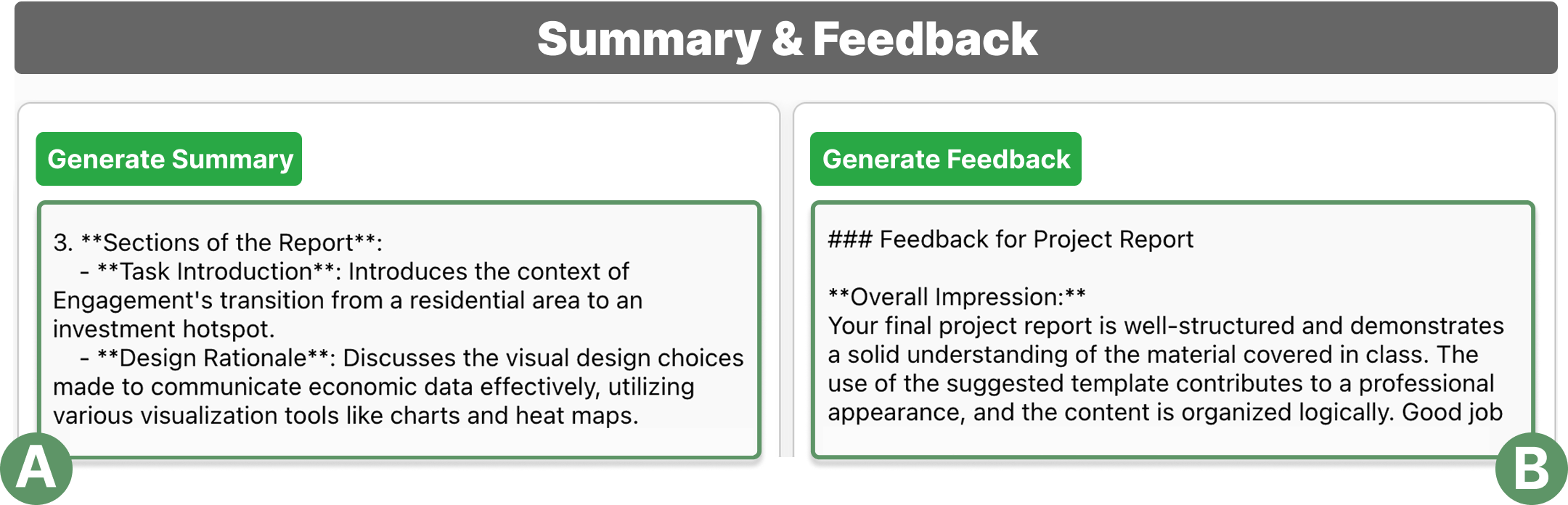}
    \caption[]{The Summary and Feedback zone, where users can view an AI-generated summary of the selected report (A). Users can also generate feedback for the report (B), which integrates highlights and comments from Focused Zone and Annotation Zone (\cref{fig:reportList}-B\&C), along with detailed metric scores and comments from Report List (\cref{fig:reportList}-A), to produce a comprehensive report.}
    \label{fig:summaryFeedback}
\end{figure}

In the \textbf{Benchmarking View}, user-uploaded reports are displayed in a list format (\cref{fig:reportList}-A). Initially, by clicking the ``Grade Reports'' button (\cref{fig:reportList}-a4) at the bottom of, backend AI components will quickly grade all reports based on the user-selected metrics, and return reference scores along with AI-generated comments to provide the automated preliminary analysis (\textbf{R3}). The average scores recommended by AI are displayed in the Report List (\cref{fig:reportList}-a2). Meanwhile, instructors can click the ``Show Details'' button to view detailed scores for each metric along with AI's comments (\cref{fig:reportList}-a1). These AI-recommended scores serve as a reference, and instructors are free to modify and update the scores (\cref{fig:reportList}-a3) for each metric (\textbf{R6}). To start a in-depth review of each report, instructors can click on each report card to focus and view the original report in the \textbf{Focused Zone} (\cref{fig:reportList}-B). Instructors are able to highlight, annotate, and edit freely within the \textbf{Annotation Zone} (\cref{fig:reportList}-C) during the grading process. All the metric scores, together with instructors' highlights and annotations will all be saved for the further feedback generation in the \textbf{Summary and Feedback Zone} (\cref{fig:summaryFeedback}). 

\begin{figure*}[!tbp]
    \centering
    \includegraphics[width=\linewidth]{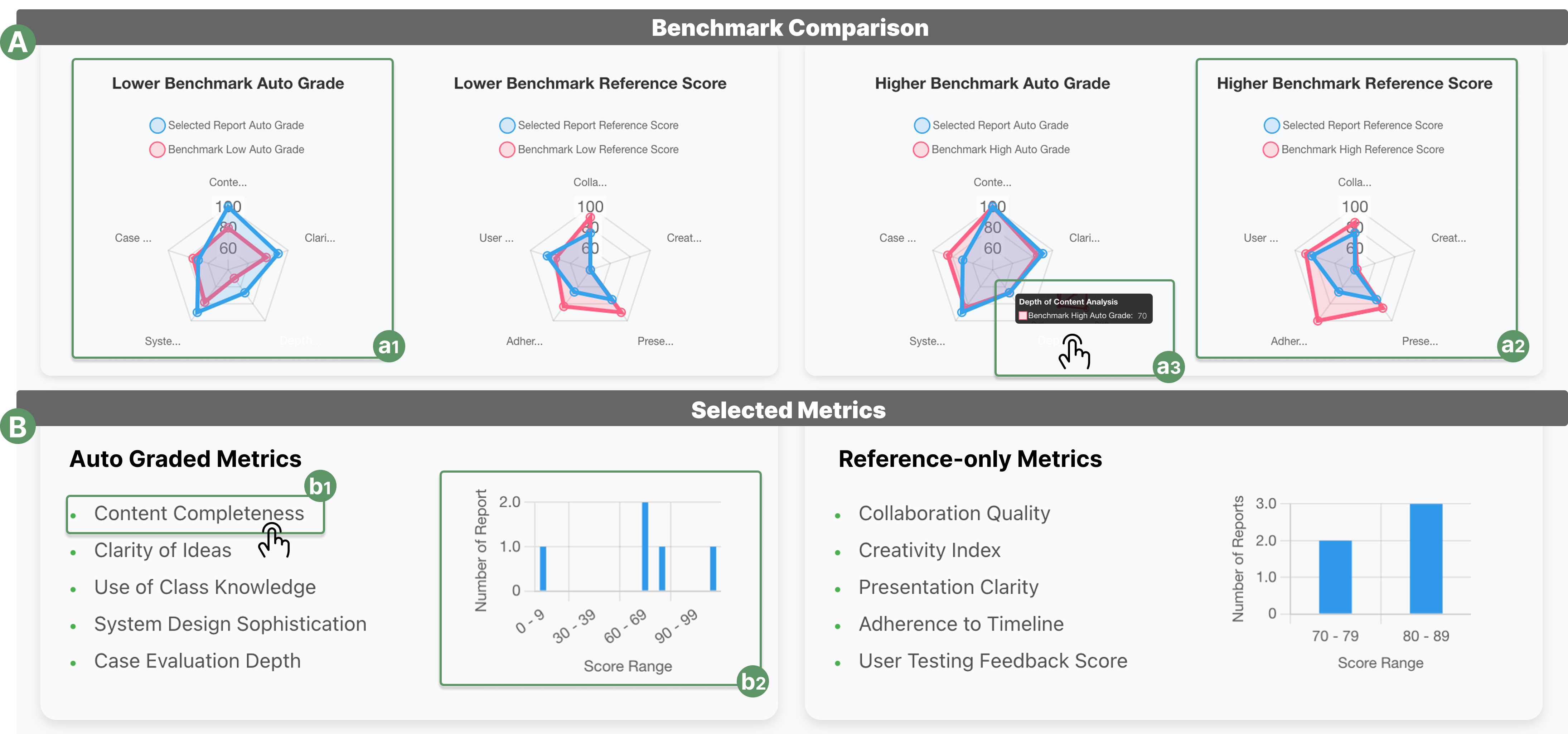}
    \caption[]{(A) The Benchmark Comparison component. Radar charts (a1 \& a2) comparing the focused report with both high and low benchmarks. Each vertex (a3) represents a user-selected metric, categorized based on user labels (auto grade or reference score) from the \textit{Metric View} (\cref{fig:metricView}-B \& C). (B) The Selected Metrics component. Users can check the selected metrics list and the corresponding score distribution charts. Users can click on each metric (b1) to view its score distribution (b2) across all reports, with the Report List (\cref{fig:reportList}-A) automatically sorted in descending order based on the selected metric.}
    \label{fig:benchmarkComparison}
\end{figure*}

During the reviewing process, once instructors identify reports suitable to serve as benchmarks, they can go back to the Report List and click the ``Set Benchmark'' button inside each report card (\cref{fig:reportList}-a1) to designate the current report as either a \textit{Benchmark High} or \textit{Benchmark Low} for further grading reference.

After instructors decide the benchmarks and select the report to focus on, their metrics scores will be updated in the \textbf{Benchmark Comparison Zone}, where users can easily compare the focused report with the selected benchmarks under each metric for a further grading adjustments (\cref{fig:benchmarkComparison}-A). The \textbf{Benchmark Comparison Zone} features four radar charts, with each vertex representing a unique metric. The first two radar charts compare the focused report with the \textit{Lower benchmark} (\cref{fig:benchmarkComparison}-a1), and the last two compare it with the \textit{Higher benchmark} (\cref{fig:benchmarkComparison}-a2). A consistent color coding scheme is used, with red representing the benchmarks and blue representing the focused report. Instructors can hover over the labels in the radar charts to view the full name of each metric (\cref{fig:benchmarkComparison}-a3). The radar charts provide a clear comparison for the focused report, helping instructors to quickly review the AI's grading for each metric. Additionally, an overview of the metric scores distribution among all reports is provided below, clustered into two categories based on the user's original selection in \textbf{Metric View} (\cref{fig:benchmarkComparison}-B).While reviewing individual metric values, users can sort the report cards in the Report List below (\cref{fig:reportList}-A) by clicking on each metric name to arrange them in descending order of scores (\cref{fig:benchmarkComparison}-b1). These functions streamlines the benchmarking process, aligning with the proposed design requirement (\textbf{R4}) in a convenient and efficient manner.


\begin{figure*}[!tbp]
    \centering
    \includegraphics[width=\linewidth]{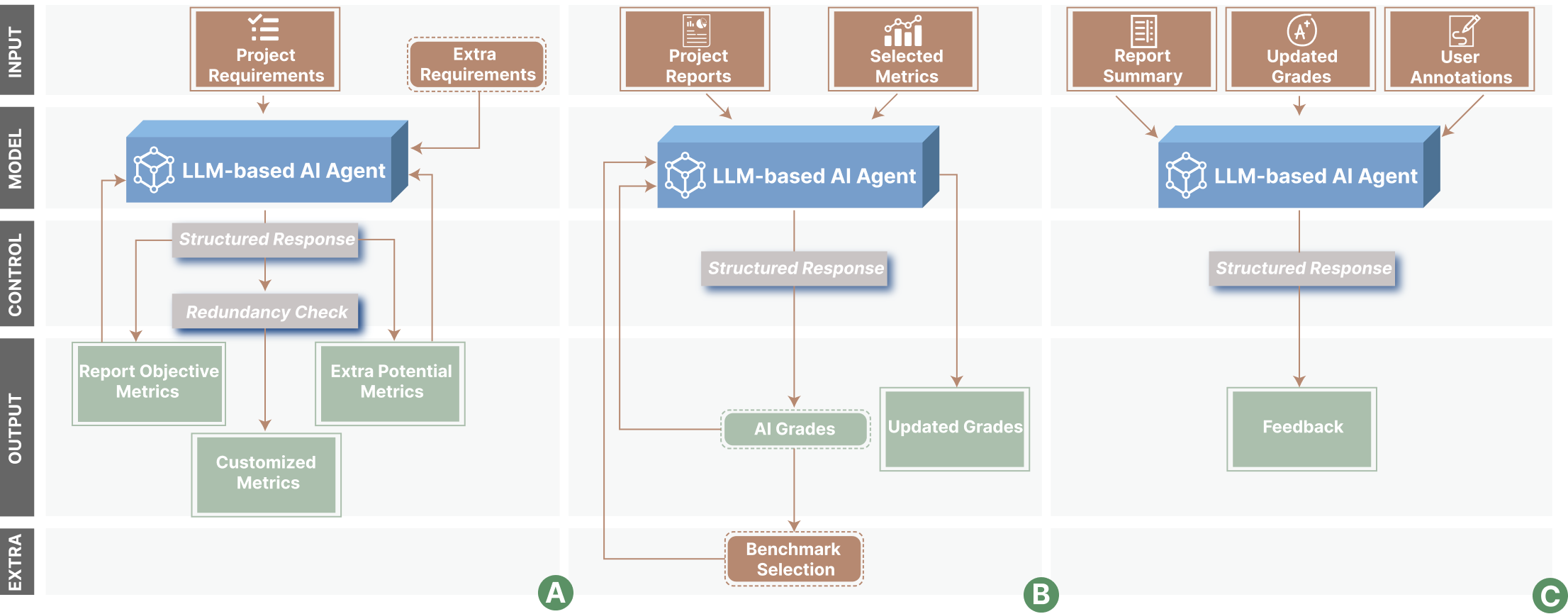}
    \caption[]{(A): In the \textit{Collaborative Metric Design} component, uploaded project requirements are automatically analyzed by an LLM-based AI agent built using the Assistant API. The output is structured into \textit{Report Objective Metrics} and \textit{Extra Potential Metrics}. Users can also input additional requirements, which the AI agent will check for redundancy before generating \textit{Customized Metrics}. (B): In the \textit{Benchmarking-driven Grading} component, both project report files and user-selected metrics are sent to another AI agent for initial grading. After users review and edit the first-round grades and select benchmark reports, the updated grades and benchmarks are sent back to the AI agent for regrading the reports. (C): In the \textit{Feedback Generation} component, The AI-generated report summary, updated grades, and user annotations are sent to the extra AI agent to generate structured feedback for students.}

    \label{fig:backend}
\end{figure*}


\revision{Throughout the AI-assisted grading process, instructors often perform multiple iterations of benchmark selection and regrading—a task that can be time-consuming and cognitively demanding when aiming to ensure fairness and consistency. To support this process, {\systemName} provides a regrade function aligned with our design goals (\textbf{R4}) and safeguards the balance of reliance on AI (\textbf{R6}). Once benchmarks are selected, instructors can trigger automated regrading by clicking the ``Regrade Reports'' button (\cref{fig:reportList}-a4), prompting the system to re-evaluate each report against the selected metrics, benchmarks, and comments. Crucially, these AI-generated updates also serve as supportive suggestions only and can be freely discarded or revised by instructors. Instructors are encouraged to critically review these outputs and repeat the regrading process as needed to ensure the outcomes align with their pedagogical standards (\textbf{R6}).}



Finally, once the benchmarking-driven grading process is complete, instructors can generate summarized feedback for each report to share with students. By clicking the ``Generate Feedback'' button in the \textbf{Summary \& Feedback} zone (\cref{fig:summaryFeedback}-A,B), AI will automatically compile all metric grades and comments, along with the instructor's highlights and annotations from (\cref{fig:reportList}-A,B,C), to generate personalized summary and feedback for each report (\cref{fig:summaryFeedback}-A,B). \revision{The feedback is structured to prioritize instructor-authored content, followed by instructor-edited AI suggestions, with pure AI-generated comments used only to fill remaining gaps (\textbf{R5}). Instructors remain free to review and revise the output before exporting it to students (\textbf{R6}). These features support consistent and personalized feedback, enhancing the student learning experience while reducing the instructor's workload (\textbf{R5 \& R6}).}



\subsection{{\systemName} Backend LLM Development} 
In this section, we introduce the backend structure of {\systemName}, focusing on the design and integration of LLMs.

The core AI functionality of {\systemName} is built using OpenAI's ChatGPT-4o API \cite{gpt2024}, which offers a lightweight yet efficient solution, striking a balance between fast response times and model performance. The backend architecture aligns with the frontend system design and is divided into three main components: \textit{Collaborative Metric Design} (\cref{fig:backend}-A), \textit{Benchmarking-driven Grading} (\cref{fig:backend}-B), and \textit{Feedback Generation} (\cref{fig:backend}-C). Each of these components leverages the \textit{Assistant API} to build AI agents that support report grading, file search, and conversation context management—ensuring a smooth and cohesive grading experience~\cite{assistant2024gpt}.

\revision{In the \textit{Collaborative Metric Design} component, the LLM-based AI agent processes users' project requirement inputs and any extra requirement descriptions (\cref{fig:backend}-A)}. To ensure structured and consistent output, the \textit{Structured Response} function is employed \cite{structured2024gpt}. The first round of requirement analysis generates two sets of metrics: \textit{Report Objective Metrics} and \textit{Extra Potential Metrics}. While users are free to specify additional metrics using natural language descriptions, all \textit{Customized Metrics} are passed through a \textit{Redundancy Check} to prevent duplication or overlap with existing metrics. After this process, the system generates three categories of metrics (with explanations and recommended formulas), allowing users to review and select appropriate ones for the grading process.

\revision{In the \textit{Benchmarking-driven Grading} component, another LLM-based AI agent processes the uploaded project reports and applies the selected metrics (\cref{fig:backend}-B)}. The system uses the \textit{File Search} tool provided by the Assistant API \cite{filesearch2024gpt}, a built-in retrieval-augmented generation (RAG) tool that enables efficient file search and content analysis \cite{chen2024benchmarking}. \revision{For the initial grading, we prompt the AI to ``think'' and ``reason'' about each report based on the selected metrics. It searches for relevant evidence within the report, assigns initial scores, and generates reference comments with supporting excerpts.} After the instructor updates the AI-generated scores and selects benchmark reports based on their expertise, the system regrades each report by comparing them against the benchmarks and adjusting the scores accordingly. \revision{During the benchmarking-based regrading stage, we prompt the AI to ``compare'' the target report and its current evaluations (scores and comments for each metric) against the selected benchmarks to refine its reasoning and update the evaluation accordingly.} Once the grading process is completed, the \textit{Updated Grades} and comments are sent back to the frontend for review.


\revision{In the \textit{Feedback Generation} component, an additional LLM-based AI agent synthesizes user-updated scores, comments, annotations, and report summaries to produce personalized feedback for each student (\cref{fig:backend}-C)}. To maintain consistent feedback style, the output is structured through the \textit{Structured Response} feature. This ensures uniformity and clarity in the feedback provided to students. Detailed prompt templates for each components are available in the supplementary materials.

\subsection{Implementation}
{\systemName} is developed as a full-stack web application, featuring a React.js \cite{fedosejev2015react} frontend for an intuitive and responsive user interface. The backend, built with Python using Flask API \cite{grinberg2018flask}, hosts and manages the LLM modules. The frontend and backend are connected via RESTful APIs, allowing smooth communication and integration between the user interface and the LLM-powered grading modules.

%% file: sections/05-evaluation.tex
\section{Evaluation}

\revision{We conducted an evaluation study to assess the effectiveness of the proposed human-AI collaborative workflow in {\systemName}, as well as the performance of each component within the system. Our primary goal was to evaluate how well the workflow enhances grading efficiency, reliance, and consistency through LLM assistance. To further evaluate concerns about over-reliance on AI, we also examined participants’ actual engagement behaviors with AI-generated outputs. To explore these objectives, we addressed the following research questions:}

    \textbf{RQ1:} How effective is the metric generation step in helping instructors define and select assessment metrics?
    
    \textbf{RQ2:} How well does the LLM’s grading and comments align with instructors' assessments, and how do instructors perceive the efficiency, reliance and quality of LLM-generated scores and comments?
    
    \textbf{RQ3:} How does the benchmark-driven grading process impact grading consistency and efficiency? 
    
    \textbf{RQ4:} How do instructors perceive the reliance of the LLM-generated feedback, and to what extent does it help streamline the feedback process?
    
    \textbf{RQ5:} What are instructors' overall attitudes toward the human-AI collaborative approach in {\systemName}, and how do they evaluate its impact on improving the grading workflow?
    
    \textbf{RQ6:} How closely do the instructors' grading results with {\systemName} align with their original, purely human-assigned report grades?

    \revision{\textbf{RQ7:} To what extent do instructors actively verify, adjust, or override AI-generated outputs during the grading process?}

\subsection{Procedure of User Study}
Participants were recruited from the authors' institution via email invitation during the summer of 2024. This study received approval from the Institutional Review Board (IRB) of the authors' institution. After obtaining participants' consent, demographic data were collected anonymously, focusing on their experience as instructors, their teaching roles, length of experience, and previous attitudes toward using LLMs like ChatGPT for grading. 

\revision{Each user study session lasted approximately 80 minutes and consisted of three stages: (1) a 50-minute system exploration and grading experiment, (2) a 15-minute questionnaire to collect participants’ perceptions of each step in the workflow, and (3) a 15-minute post-interview to gather qualitative feedback on user experience and system design.}

Before beginning the grading experiment, participants were provided with a brief background introduction and a system demo. They were shown how to use the proposed workflow, covering metric generation, benchmark-driven grading, and feedback generation steps. Participants were provided with a sample project requirement and five sample project reports from a real-world master-level data visualization course conducted in the previous semester at the authors' institution. These materials were collected with IRB approval and consent from both the instructors and the student authors. Participants were required to upload both the project requirements and reports to the system and begin the grading process. Their primary task was to provide a final ranking, assign reference scores (out of 100), and determine reference letter grades (from F to A) for each report. A score of 85 and above corresponds to an A, with every 5-point interval representing a subgrade (e.g., 80-85 is A-, 75-80 is B+, and so on) \revision{To support the analysis of instructors' engagement with AI-generated outputs, we logged participants' interactions with the system, including the number of selected metrics, the frequency of manual comment additions or edits, and the proportion of AI-generated content that was adjusted or overridden. The experimental results and logs were later used in several quantitative analyses to evaluate the system’s effectiveness (\textbf{R6}) and instructors’ level of engagement with AI assistance (\textbf{R7}). }



\revision{After completing the grading experiment, participants were given a 7-point Likert-scale questionnaire consisting of multiple rating questions to collect their opinions on each step of the workflow, following the research questions (RQ1–RQ5)}. A grading scale of 1 to 7 was used to represent participants' negative to positive attitudes toward each feature. Finally, after completing the questionnaire, participants were invited to a brief interview to share their thoughts and comments on each step of the workflow and the system itself.

\subsection{Project Report Data Collection}
The project report requirements, submissions, as well as the ground truth final scores and letter grades, were collected from a master-level data visualization course at the authors' institution during the spring semester of 2024. This project served as the main assessment for the course, in which each group of students was asked to choose an open dataset, identify specific research questions of interest, and apply the knowledge they gained throughout the course to analyze the dataset using visual analytics techniques. The students reported their methods and findings in a formal written report. The final scores were determined by taking a weighted average of evaluations from one course instructor and two senior teaching assistants. We obtained consent from both the course instructors and students to use their reports for our research at the end of the semester, and we also received IRB approval for the data collection. For the 5 sample reports used in the user experiments, we selected reports with letter grades ranging from B- to A, aiming to reflect the real distribution of students' work. All student personal information was anonymized before being used in the experiments to ensure privacy and confidentiality.

\subsection{Participants}
We recruited 12 participants (P1-P12, 8 male and 4 female) through email invitations to instructors and teaching assistants at our institution who had experience teaching this data visualization course. Of the participants, four (P1-P4) were university professors or lecturers, and eight (P5-P12) were senior PhD students with teaching assistant experience in this course. On average, participants had 3.25 years of experience in designing or grading project reports. Among the 12 participants, seven had experience using LLMs to assist with grading, while five did not. Each participant spent approximately one hour on the study and received \$30 in compensation.

\subsection{User Study Results}

In this section, we discuss our findings regarding instructors' experiences with {\systemName}, focusing on their perceptions of each step in the proposed workflow: collaborative criteria generation, the benchmark-driven grading process, and feedback generation.


\begin{figure}[!tbp]
    \centering
    \includegraphics[width=\linewidth]{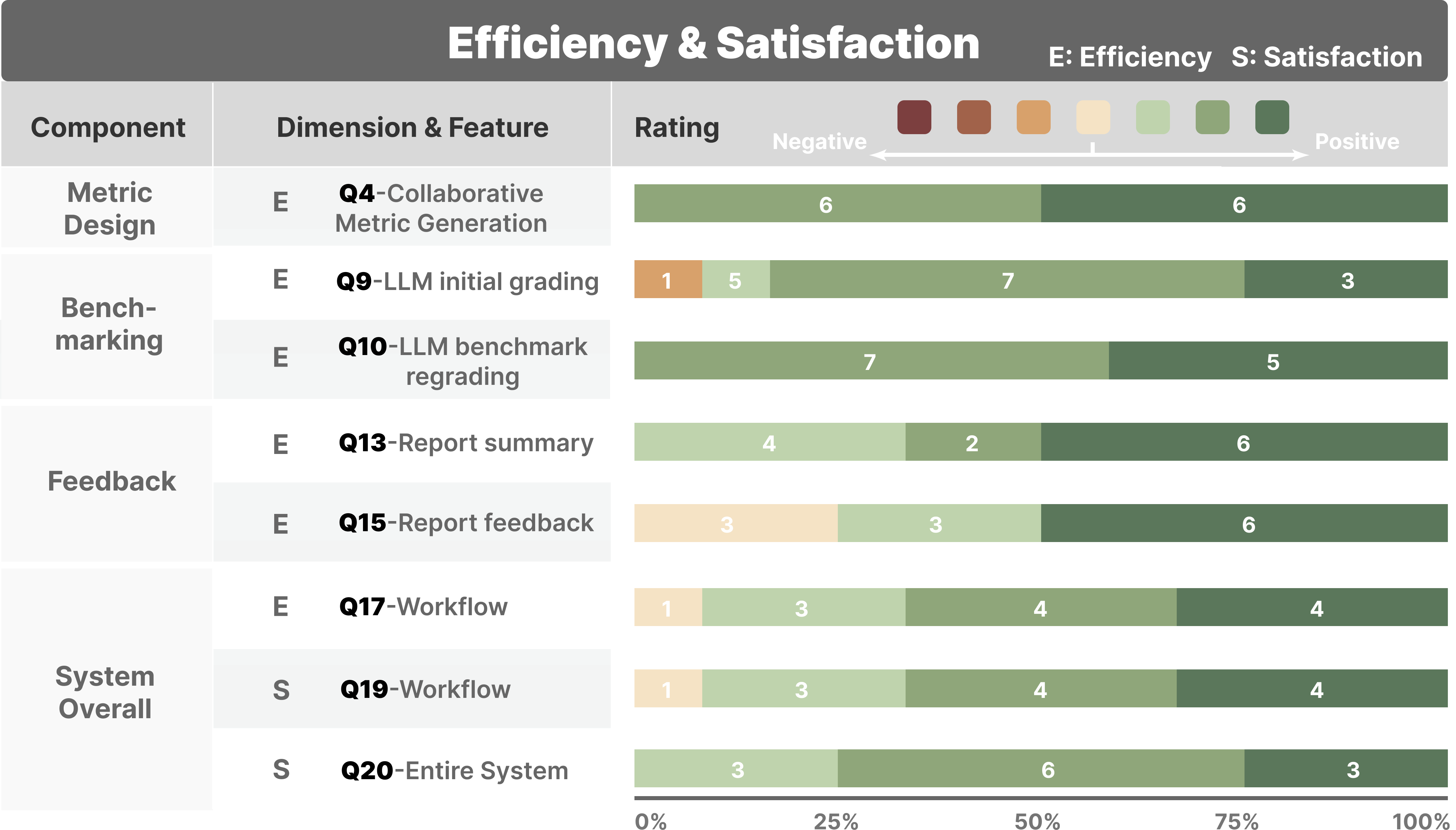}
\caption{Summary of user study results on participants' ratings for each system component's Efficiency and Satisfaction. Each question was rated by participants on a 7-point scale, where dark brown represents the strongest negative response, dark green indicates the strongest positive response, and neutral beige marks the midpoint. The number inside each colored bar reflects the count of participants who selected that specific rating.}
    \label{fig:userStudy_2}
\end{figure}


\subsubsection{RQ1: Instructors find the collaborative metric generation step efficient, effective, and acceptably reliable.}

To evaluate instructors' experiences with the collaborative metric design step, we asked participants to rate the efficiency and reliability of various metric generation methods on a 7-point scale, with 4 representing neutrality. Overall, participants rated the efficiency of the collaborative metric generation process very positively (\textbf{Q4:} mean = 6.5, stdev = 0.522, ~\cref{fig:userStudy_2}). For the reliability of each metric design component, the \textit{Report Objective Metrics} feature (automatic analysis of report requirements) received positive ratings (\textbf{Q1:} mean = 5.91, stdev = 0.793, ~\cref{fig:userStudy_1}), highlighting the strong ability of LLM to convert the requirements of the report into concrete quantitative metrics. Similarly, the \textit{Extra Potential Metrics} feature (automatic metric recommendation beyond report requirement) was highly rated (\textbf{Q2:} mean = 6.16, stdev = 0.937, ~\cref{fig:userStudy_1}), with participants appreciating its ability to brainstorm additional useful metrics with clear explanations. On the other hand, the \textit{Custom Metric Design} feature (metric co-design through user inputs) also scored well for its reliability (\textbf{Q3:} mean = 5.83, stdev = 1.115, ~\cref{fig:userStudy_1}). Participants overall expressed satisfaction with the collaborative metric design step. However, participant P4 noted occasional misunderstandings by the LLM when using the \textit{Custom Metric Design} function and suggested allowing users to freely edit the metric explanations and formulas to improve the process. Participants also recommended adding features like semantic grouping and metric weighting for further customization.

In general, participants felt that the collaborative metric generation step aligned well with their workflow, significantly enhancing both grading efficiency and reliability.

\subsubsection{RQ2: Instructors find the LLM's grading and comments efficient, with generally acceptable reliability and alignment with their assessments.}

\begin{figure}[!bp]
    \centering
    \includegraphics[width=\linewidth]{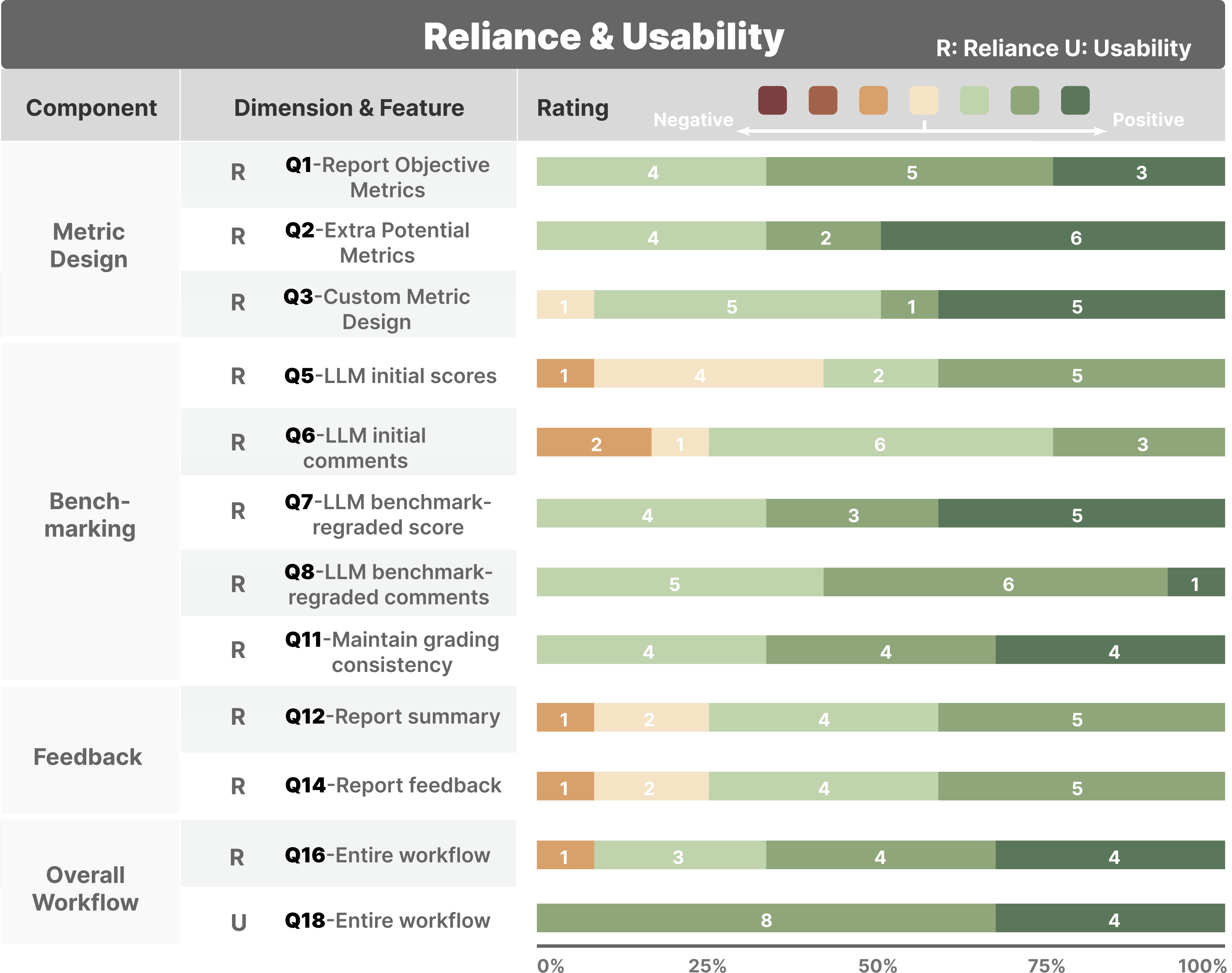}
\caption{Summary of user study results on participants' ratings for each system component's Reliance and Usability. Each question was rated by participants on a 7-point scale, where dark brown represents the strongest negative response, dark green indicates the strongest positive response, and neutral beige marks the midpoint. The number inside each colored bar reflects the count of participants who selected that specific rating.}
    \label{fig:userStudy_1}
\end{figure}

Regarding the efficiency of LLM-generated scores and comments, participants expressed strong positive attitudes (\textbf{Q9:} mean = 5.92, stdev = 1.083, ~\cref{fig:userStudy_1}) towards the LLM's initial grading, and even more favorable attitudes (\textbf{Q10:} mean = 6.42, stdev = 0.515, ~\cref{fig:userStudy_1}) towards the benchmark regrading. Participants noted the system’s ability to handle multiple metrics and compare reports with quick response times as particularly valuable.

In terms of the reliability of LLM-generated scores and comments, participants rated the benchmark-regrading process highly, with benchmark-regraded scores (\textbf{Q7}) receiving a mean of 6.08 (stdev = 0.900, ~\cref{fig:userStudy_1}) and comments  (\textbf{Q8}) receiving a mean of 5.83 (stdev = 1.11, ~\cref{fig:userStudy_1}). However, the reliability of the LLM's initial scores (\textbf{Q5:} mean = 4.91, stdev = 1.083, ~\cref{fig:userStudy_1}) and initial comments (\textbf{Q6:} mean = 4.83, stdev = 1.030, ~\cref{fig:userStudy_1}) were rated slightly lower. One participant (P11) noted that the comments were sometimes too generic, applying to multiple reports, which reduced their specificity. P8 and P11 also mentioned that certain selected metrics were occasionally missed in the initial grading and suggested adding an automatic checking function to help refine the LLM's initial output. The user editing function was considered crucial for refining the initial grading by allowing adjustments to the LLM's outputs.

From the interviews, the majority of participants found the scores and comments to be reliable and contextually appropriate. Participants particularly appreciated the benchmark-regrading process, with P12 stating, \textit{``The regrading process specifically compares each report with the benchmarks I selected. I can clearly see from the comments why one report performed better or worse than the benchmarks under each metric, which is very helpful and significant. I also found that most of the LLM comments aligned with my assessment.''}


\subsubsection{RQ3: Instructors find the benchmark-driven grading process significantly improves grading consistency and efficiency.}

Participants gave highly positive ratings for the benchmark-driven regrading process, praising both its efficiency (\textbf{Q10:} mean = 6.42, stdev = 0.515, ~\cref{fig:userStudy_2}) and reliability (\textbf{Q11:} mean = 6.00, stdev = 0.853, ~\cref{fig:userStudy_1}). They found that this process significantly reduced their workload and maintained grading consistency. As P1 remarked, \textit{``The system excels in the benchmarking process. AI makes grading faster, and the comparison charts simplify the process. It reduced my effort significantly, allowing me to focus more on content quality and ensure my judgment is more unbiased''}. Participants (P1, P2, P5 \& P10) also appreciated that most comments included specific comparisons with the benchmark reports after the LLM regrading. Beyond the LLM's role, the clear radar charts also provided visual comparisons and references, making it easier for participants to make grading decisions. On the other hand, participants valued the ability to select and edit benchmarks. As P9 noted, \textit{``For the second round of grading, it performed much better. I think the benchmarks I set made a difference. The metric scores and comments I edited in the benchmarks were definitely important. It's good to have control, especially over key aspects like benchmark references.''}

Overall, participants expressed trust and satisfaction with the LLM’s performance in the benchmarking process, particularly in regrading and comparing reports. The combination of LLM-powered analysis and instructor-driven benchmark selection proved to be a powerful tool for improving grading consistency and efficiency.

\subsubsection{RQ4: Instructors find the quality of the AI-generated summary and feedback helpful in streamlining the feedback process.}


Participants gave strong positive ratings for the efficiency of the AI-generated summary (\textbf{Q13:} mean = 6.33, stdev = 0.778, ~\cref{fig:userStudy_2}) and feedback (\textbf{Q15:} mean = 6.00, stdev = 1.279, ~\cref{fig:userStudy_2}), along with positive ratings for reliability (\textbf{Q12 \& Q14:} mean = 5.08, stdev = 0.996, ~\cref{fig:userStudy_1}) for both processes. Participants appreciated how this simple yet effective feature provided a quick summary of the students' reports, allowing them to quickly grasp the topic and content. They also found the feedback function particularly useful. As P10 mentioned, \textit{``The feedback function is very helpful, as it not only collects the metrics, scores, and comments I edited but also includes the annotations and highlights I made. In my past experience, I often felt exhausted after writing feedback at the end of a long grading process, especially when I had many bullet points to cover''.} However, P3 noted that while the AI-generated feedback is a good reference, it should not be used directly when providing feedback to students, as it could appear too generic and overly ``AI-styled'', making students feel their hard work wasn't given proper attention.

\begin{table*}[!t]
\tiny

\resizebox{\linewidth}{!}{
\begin{tabular}{p{2.4cm}|ccccc}
\toprule
\textbf{Statistics} & \multicolumn{5}{@{\hspace{-0.5em}}c}{\textbf{Report ID}} \\
                    & R01 & R02 & R03 & R04 & R05 \\
\midrule
Avg.score (Ps) & 90.9 & 75.0 & 74.5 & 84.1 & 83.3 \\
Ground truth score & 89.1 & 75.6 & 71.5 & 79.3 & 82.9 \\
\midrule
Assigned Grades (Ps) & 9A 2A- 1B+ & 1A- 1B+ 5B 5B- & 1A- 2B+ 3B 4B- 2C & 3A 2A- 5B+ 2B & 2A 2A- 7B+ 1B- \\
Ground truth grade & A & B & B- & B+ & A- \\
\bottomrule
\end{tabular}
}
\caption{The summary of grading consistency analysis: (1) Comparison of participants’ average scores for each report with the ground truth scores. (2) Comparison of participants' assigned letter grades for each report with the curved ground truth grades.}
\label{tab:quantitativeTest}
\end{table*}

\subsubsection{RQ5: Instructors overall have a very positive impression of the human-AI collaborative workflow in {\systemName}, finding it greatly supports the grading process.}


Overall, participants expressed strong positive attitudes toward the workflow and system usability. Specifically, the workflow's efficiency was well-regarded (\textbf{Q17 \& Q19:} mean = 5.92, stdev = 0.996, ~\cref{fig:userStudy_1,fig:userStudy_2}), with good ratings for workflow reliability (\textbf{Q16:} mean = 5.83, stdev = 1.193, ~\cref{fig:userStudy_1}). The system was also rated as very easy to use (\textbf{Q18:} mean = 6.33, stdev = 0.492, ~\cref{fig:userStudy_1}), and overall satisfaction with the system was high (\textbf{Q20:} mean = 6.00, stdev = 0.739, ~\cref{fig:userStudy_2}). In summary, our user study and analysis suggest that the proposed collaborative workflow and system are helpful, effective, and reliable in most cases. However, some challenges, particularly in LLM stability and ethical considerations, remain to be addressed. In the Discussion section, we will delve deeper into instructors' perspectives on this human-AI collaborative grading process, potential ethical issues, and design considerations for similar tools.



\subsection{Quantitative Analysis of Grading Consistency}

\revision{To evaluate the internal consistency of instructor grading when using {\systemName}, we analyzed the alignment between participants’ grading score distribution and the ground truth grades provided by the course instructor. Ensuring internal consistency is crucial for maintaining fairness and reducing biases introduced by the system—concerns highlighted by instructors during formative studies.}

\revision{To this end, we employed three widely used statistical metrics: Kendall’s Tau \cite{arndt1999correlating}, Spearman’s Rank Correlation \cite{sedgwick2014spearman}, and Pearson Correlation \cite{sedgwick2012pearson}. These metrics collectively assess consistency from both ranking and scoring perspectives, providing a robust evaluation of whether the distribution of participant-assigned grades aligns with the ground truth. The results for Kendall’s Tau, Spearman’s Rank Correlation, and Pearson Correlation were 0.799, 0.899, and 0.900, respectively. These strong correlations indicate that the distribution of grades assigned by each participant closely aligns with the ground truth, confirming low variation and strong internal consistency within individual graders.}



\subsubsection{Consistency in Letter Grades}

\revision{In addition to evaluating internal consistency within each grader, we also examined inter-grader consistency, as instructors highlighted fairness and reducing inter-grader bias also as key priorities. Specifically, we analyzed the consistency of letter grades assigned by different participants against the ground truth, as letter grades are often the formal outcomes communicated to students and central to academic decisions. Assessing their consistency provides practical validation of the system’s reliance. For example, in the case of report 01, the 12 participants assigned 9 A’s, 2 A-’s, and 1 B+, closely reflecting the ground truth score of ``A''. A detailed comparison between the letter grades assigned by different participants and the ground truth is shown in ~\Cref{tab:quantitativeTest}. This consistent alignment across letter grades further demonstrates the system’s robustness and reliability in supporting fair assessment of diverse project reports.}

The findings from our evaluation study suggest that {\systemName} not only aligns well with traditional grading methods but also paves the way for more efficient, scalable, and reliable grading practices in PBL environments.

\subsection{Quantitative Analysis of Instructor Engagement with AI-Generated Outputs}

\revision{To examine the extent to which instructors critically engaged with AI-generated content \textbf{(RQ7)}, we conducted a quantitative analysis of participants' interaction logs during the grading process. Specifically, we analyzed three key behaviors: metric selection, manual comment contributions, and modification of AI-generated feedback.}

\revision{Participants selected an average of 4.50 Automatic Metrics and 5.83 Reference Metrics per report, indicating a balanced use of system-generated assessment criteria. In addition, they manually added an average of 5.25 personalized comments per report, demonstrating that instructors actively contributed their own evaluations rather than relying solely on AI suggestions.}

\revision{In terms of verification behavior, 9 out of 12 participants explicitly reported manually checking most AI-referenced facts in the Automatic Metrics against the original student reports. The remaining three manually verified high-impact criteria such as report structure or alignment with project requirements, while skipping low-stakes aspects like grammar or formatting.}

\revision{When using Reference Metrics, instructors adjusted or overrode an average of 64.09\% of the AI-generated scores and comments. All participants reported that they reviewed nearly every AI-suggested comment before making a decision to accept, modify, or discard it.}

\revision{These findings confirm that instructors engaged with AI-generated outputs in a deliberate and selective manner. Rather than relying on AI passively, participants exercised consistent judgment in validating and revising AI assistance—highlighting the effectiveness of our system design in promoting instructor authority and supporting balanced use.}

%% file: sections/06-discussion.tex
\section{Discussion}
In this paper, we explored the potential of LLMs to assist instructors in grading project reports and developed a tool aimed at supporting this critical learning assessment process. 
In the following sections, we will discuss how our study contributes to existing research on human-AI collaboration in educational assessment tools. Specifically, we will explore instructors' perspectives, ethical concerns, and the design considerations and implications that can guide the future development of such systems. We will conclude by addressing the limitations of our study and outlining potential directions for future research.




\subsection{Instructor Perspectives on Leveraging LLMs for Grading}

Our study revealed that instructors generally appreciated the integration of LLMs to streamline the grading process. Participants, such as P6 and P8, emphasized how the use of LLMs helped them brainstorm potential metrics, gain external perspectives on grading, reduce the back-and-forth effort during the grading process, and organize their thoughts to provide more structured feedback. Overall, most participants expressed a strong positive attitude toward using LLMs to assist in the grading process. 

However, concerns were raised about the reliability of certain parts of the LLM-generated outputs. For instance, P6, P9, and P11 observed that LLMs sometimes overlooked specific grading metrics or neglected certain report sections. Due to these inherent limitations, it is challenging to guarantee that AI outputs are consistently thorough and comprehensive. Introducing safeguards or providing more flexibility for human calibration could address these shortcomings. As P9 suggested, enabling iterative refinement through a dialogue with the LLM could allow instructors to guide the system in regrading specific sections without a complete reevaluation, improving both accuracy and usability. Ultimately, such enhancements could help foster a more balanced and reliable integration of AI in the grading process.


\subsection{Social Risks and Pedagogical Implications of LLM-Assisted Grading}

\revision{While LLMs offer clear potential to support grading workflows, their integration also raises important social and pedagogical concerns—particularly the risk of eroding personalized feedback, diminishing instructor engagement, and reducing student trust. These concerns reflect broader challenges in maintaining human presence and pedagogical authenticity in AI-assisted grading, and point to the importance of understanding students’ perceptions.}

\revision{A key insight from our study is that AI-supported grading systems should be explicitly designed to enhance—rather than diminish—personalized and actionable feedback. As participant P11 noted, one of the \systemName's most valuable features was the benchmark-driven workflow, which supported instructors in crafting more thoughtful and targeted feedback. By enabling comparisons with exemplar reports on selected metrics—supplemented by AI-generated scores, comments, and contextual evidence—{\systemName} provided a structured yet flexible framework for evaluating student work. This design allowed instructors to offload routine tasks and focus more deeply on higher-order, expertise-intensive evaluation—leveraging AI to strengthen, rather than compromise, the quality of feedback and instruction.}

\revision{Another important design consideration is the structure and transparency of AI-assisted feedback. To maintain student trust and pedagogical authenticity, instructors emphasized that human-authored comments should remain central and be clearly distinguishable from AI-generated content. In our study, participants responded positively to \systemName’s feedback generation, which prioritized instructor-written comments, followed by instructor-edited AI suggestions, and finally, pure AI-generated comments used primarily to fill remaining gaps. Future systems may benefit from incorporating more explicit visual or textual cues (e.g., labels or icons) to indicate the origin of each comment—especially in high-stakes or formative assessment contexts where transparency and trust are essential.}

\revision{Students’ perspectives further reinforced the importance of thoughtful feedback design in AI-assisted grading. In our informal discussions with students from the course that provided the project report data, many expressed openness to the use of AI, noting its potential to reduce overlooked content, mitigate instructor bias toward certain topics, and accelerate feedback turnaround. At the same time, several students mentioned they could often ``tell when the feedback wasn’t really written by the teacher'', particularly when it felt overly generic or formulaic. These insights highlight both the perceived benefits and the importance of addressing the social risks of AI involvement in assessment design—particularly the need to preserve authenticity and transparency to foster trust and support constructive engagement with feedback.}

\subsection{Design Considerations for calibration of LLM's grading accuracy}
A recurring theme in the interviews was the need for more nuanced calibration of LLM grading accuracy. Participants such as P2, P5, and P11 observed that, while benchmarking generally aligned with their grading habits, the system occasionally showed inconsistencies or missed certain grading elements. To address this, participants proposed a multi-level calibration process involving human oversight at the metric, section, and overall report levels.

At the metric level, a targeted regrade function could allow instructors to select specific metrics for re-evaluation. Before initiating regrading, instructors could interact with the LLM, providing comments, restrictions, or instructions to refine its focus on particular metrics.

For section-level calibration, enabling granular grading within specific report sections would offer instructors greater control. As suggested by P11, this approach could improve consistency across submissions by allowing instructors to fine-tune the grading for distinct report sections.

At the overall report level, incorporating a chatbot into the system could facilitate real-time interaction, allowing instructors to pose questions or clarify discrepancies in the LLM’s grading approach. This ``discussion'' feature would be especially valuable during benchmarking, helping to reconcile differences in judgment and reduce personal biases, ultimately supporting greater fairness and accuracy in grading.

\subsection{Design Implications for Instructor-AI Collaborative Grading System}

Several design implications emerged from our study on how to enhance instructor-AI collaboration in grading. One key takeaway was the importance of integrating multiple functional components into a single, interconnected interface. Instructors appreciated this seamless experience, as highlighted by P10, who valued the integration of comments, PDF annotation, AI grading references, and AI-generated feedback within a unified interface. This streamlined workflow allowed her to avoid toggling between different platforms, enhancing efficiency and reducing cognitive load.

Additionally, P1 and P9 emphasized the importance of allowing instructors to first review the report summary before interacting with LLM-generated scores and comments. This step enables instructors to form their own independent impression of the report, helping to mitigate potential bias introduced by the AI assessments. Such an approach would be instrumental in preserving instructor autonomy and ensuring that the LLM outputs are used as a reference, rather than dictating the grading process. P9 further suggested leveraging the design of benchmarking radar charts to provide more informative and personalized feedback to students. By including personalized comparisons against ``role model'' reports across multiple metrics, students could gain clearer insights into how their work aligns with established standards. Integrating visual aids like radar charts into feedback would offer a transparent and easily interpretable representation of performance relative to benchmarks, making feedback more instructive and actionable.


Beyond ensuring trustworthy grading accuracy, {\systemName} provides substantial benefits in grading efficiency, a feature that was highly appreciated by the instructors. By automating the generation of quantitative metrics, handling initial evaluations, and enabling benchmark-driven regrading, the system reduces the cognitive effort required from instructors. This allows them to focus more on qualitative assessments that necessitate human expertise. It also frees up time for instructors to engage in more personalized instruction and intervention. 

Moreover, the involvement of AI in grading can help mitigate personal biases, ensuring a more consistent and fair evaluation process. This not only addresses the issue of limited instructor and teaching assistant availability but also improves the fairness of grading. As educational demands continue to grow, AI-driven tools like {\systemName} have the potential to play a pivotal role in enhancing assessment workflows, making grading more efficient and equitable.

\subsection{Limitations and Future Work}
While our system and workflow demonstrate strong potential, several limitations emerged during the study. One notable issue, as raised by P11, is the inconsistency of LLM-generated grading—some variations in scores were observed across different attempts. This underscores the need for further research to enhance the stability of LLM outputs, particularly in subjective or complex assessment contexts. Future work could explore more robust human-AI collaboration strategies, such as iterative feedback loops or multi-step calibration processes, to reduce such inconsistencies and improve grading reliability. Secondly, as P9 pointed out, the connection between the original report content and LLM-generated comments could be more explicit, enabling automatic scrolling to the relevant text excerpts for easier feedback verification. Strengthening this link would not only improve transparency but also increase instructors’ confidence in the system's outputs. Thirdly, improving the system’s ability to handle non-textual components, such as visual elements—is also helpful, particularly in project-based learning contexts where these components are integral to the evaluation. 

\revision{Another current limitation is that the system does not yet incorporate external materials, such as code files, into its evaluation process. This limits its ability to verify certain types of factual content that rely on supplementary materials beyond the report itself. Incorporating external materials is part of our planned future work, which will expand the system’s capacity to support more comprehensive and accurate assessments, particularly in technical domains where code, datasets, and other artifacts play a central role.}

Instructors also pointed to other research opportunities, noting that the tool’s success in data visualization courses indicates strong potential for adaptation to other disciplines—particularly those with unique grading rubrics or subjective evaluation criteria, such as business presentations or design projects. Future work could further refine the system to accommodate a wider range of assessment types, enhancing its versatility and applicability across diverse educational contexts.


%% file: sections/07-conclusion.tex
\section{Conclusion}

This paper tackled the challenges of grading project reports by integrating large language models (LLMs) into the assessment process. Based on insights from our formative study, we developed a novel workflow combining instructor-AI metric co-design, benchmarking, and AI-assisted feedback. The {\systemName} tool implements this workflow to improve grading efficiency and consistency, while ensuring instructors retain control over all judgments. Our evaluation demonstrated that this approach effectively streamlines the grading process, provides valuable support to instructors, and offers students peer-comparative feedback. Future work should explore more fine-grained human calibration of LLM outputs, while maintaining an ethical balance where AI enhances—rather than replaces—human judgment. Our work contributes to advancing human-AI collaboration in educational assessment, offering a scalable approach to the growing complexity of project-based learning, and finally supporting students' learning and improvement.

%% file: sections/09_ack.tex
\begin{acks}
This work is supported by RGC GRF grant 16218724.
\end{acks}

%% file: sample-manuscript.bbl

\begin{thebibliography}{96}


\ifx \showCODEN    \undefined \def \showCODEN     #1{\unskip}     \fi
\ifx \showDOI      \undefined \def \showDOI       #1{#1}\fi
\ifx \showISBNx    \undefined \def \showISBNx     #1{\unskip}     \fi
\ifx \showISBNxiii \undefined \def \showISBNxiii  #1{\unskip}     \fi
\ifx \showISSN     \undefined \def \showISSN      #1{\unskip}     \fi
\ifx \showLCCN     \undefined \def \showLCCN      #1{\unskip}     \fi
\ifx \shownote     \undefined \def \shownote      #1{#1}          \fi
\ifx \showarticletitle \undefined \def \showarticletitle #1{#1}   \fi
\ifx \showURL      \undefined \def \showURL       {\relax}        \fi
\providecommand\bibfield[2]{#2}
\providecommand\bibinfo[2]{#2}
\providecommand\natexlab[1]{#1}
\providecommand\showeprint[2][]{arXiv:#2}

\bibitem[Abd-Alrazaq et~al\mbox{.}(2023)]%
        {abd2023large}
\bibfield{author}{\bibinfo{person}{Alaa Abd-Alrazaq}, \bibinfo{person}{Rawan AlSaad}, \bibinfo{person}{Dari Alhuwail}, \bibinfo{person}{Arfan Ahmed}, \bibinfo{person}{Padraig~Mark Healy}, \bibinfo{person}{Syed Latifi}, \bibinfo{person}{Sarah Aziz}, \bibinfo{person}{Rafat Damseh}, \bibinfo{person}{Sadam~Alabed Alrazak}, \bibinfo{person}{Javaid Sheikh}, {et~al\mbox{.}}} \bibinfo{year}{2023}\natexlab{}.
\newblock \showarticletitle{Large language models in medical education: opportunities, challenges, and future directions}.
\newblock \bibinfo{journal}{\emph{JMIR Medical Education}} \bibinfo{volume}{9}, \bibinfo{number}{1} (\bibinfo{year}{2023}), \bibinfo{pages}{148--162}.
\newblock


\bibitem[Alasadi and Baiz(2023)]%
        {alasadi2023generative}
\bibfield{author}{\bibinfo{person}{Eman~A Alasadi} {and} \bibinfo{person}{Carlos~R Baiz}.} \bibinfo{year}{2023}\natexlab{}.
\newblock \showarticletitle{Generative AI in education and research: Opportunities, concerns, and solutions}.
\newblock \bibinfo{journal}{\emph{Journal of Chemical Education}} \bibinfo{volume}{100}, \bibinfo{number}{8} (\bibinfo{year}{2023}), \bibinfo{pages}{2965--2971}.
\newblock


\bibitem[Albar and Southcott(2021)]%
        {albar2021problem}
\bibfield{author}{\bibinfo{person}{Sharifah~Balkish Albar} {and} \bibinfo{person}{Jane~Elizabeth Southcott}.} \bibinfo{year}{2021}\natexlab{}.
\newblock \showarticletitle{Problem and project-based learning through an investigation lesson: Significant gains in creative thinking behaviour within the Australian foundation (preparatory) classroom}.
\newblock \bibinfo{journal}{\emph{Thinking Skills and Creativity}}  \bibinfo{volume}{41} (\bibinfo{year}{2021}), \bibinfo{pages}{100--118}.
\newblock


\bibitem[Alves et~al\mbox{.}(2009)]%
        {alves2009teachers}
\bibfield{author}{\bibinfo{person}{Anabela~C Alves}, \bibinfo{person}{Francisco Moreira}, \bibinfo{person}{Rui~M Sousa}, {and} \bibinfo{person}{Rui~M Lima}.} \bibinfo{year}{2009}\natexlab{}.
\newblock \showarticletitle{Teachers’ workload in a project-led engineering education approach}. In \bibinfo{booktitle}{\emph{International Symposium on Innovation and Assessment of Engineering Curricula}}, Vol.~\bibinfo{volume}{14}.
\newblock


\bibitem[Arawjo et~al\mbox{.}(2017)]%
        {arawjo2017typetalker}
\bibfield{author}{\bibinfo{person}{Ian Arawjo}, \bibinfo{person}{Dongwook Yoon}, {and} \bibinfo{person}{Fran{\c{c}}ois Guimbreti{\`e}re}.} \bibinfo{year}{2017}\natexlab{}.
\newblock \showarticletitle{Typetalker: A speech synthesis-based multi-modal commenting system}. In \bibinfo{booktitle}{\emph{Proceedings of the 2017 ACM Conference on Computer Supported Cooperative Work and Social Computing}}. \bibinfo{pages}{1970--1981}.
\newblock


\bibitem[Arndt et~al\mbox{.}(1999)]%
        {arndt1999correlating}
\bibfield{author}{\bibinfo{person}{Stephan Arndt}, \bibinfo{person}{Carolyn Turvey}, {and} \bibinfo{person}{Nancy~C Andreasen}.} \bibinfo{year}{1999}\natexlab{}.
\newblock \showarticletitle{Correlating and predicting psychiatric symptom ratings: Spearmans r versus Kendalls tau correlation}.
\newblock \bibinfo{journal}{\emph{Journal of psychiatric research}} \bibinfo{volume}{33}, \bibinfo{number}{2} (\bibinfo{year}{1999}), \bibinfo{pages}{97--104}.
\newblock


\bibitem[Atlas(2023)]%
        {atlas2023chatgpt}
\bibfield{author}{\bibinfo{person}{Stephen Atlas}.} \bibinfo{year}{2023}\natexlab{}.
\newblock \showarticletitle{ChatGPT for higher education and professional development: A guide to conversational AI}.
\newblock  (\bibinfo{year}{2023}).
\newblock


\bibitem[Bahroun et~al\mbox{.}(2023)]%
        {bahroun2023transforming}
\bibfield{author}{\bibinfo{person}{Zied Bahroun}, \bibinfo{person}{Chiraz Anane}, \bibinfo{person}{Vian Ahmed}, {and} \bibinfo{person}{Andrew Zacca}.} \bibinfo{year}{2023}\natexlab{}.
\newblock \showarticletitle{Transforming education: A comprehensive review of generative artificial intelligence in educational settings through bibliometric and content analysis}.
\newblock \bibinfo{journal}{\emph{Sustainability}} \bibinfo{volume}{15}, \bibinfo{number}{17} (\bibinfo{year}{2023}), \bibinfo{pages}{129--143}.
\newblock


\bibitem[Bayerlein(2014)]%
        {bayerlein2014students}
\bibfield{author}{\bibinfo{person}{Leopold Bayerlein}.} \bibinfo{year}{2014}\natexlab{}.
\newblock \showarticletitle{Students’ feedback preferences: how do students react to timely and automatically generated assessment feedback?}
\newblock \bibinfo{journal}{\emph{Assessment \& Evaluation in Higher Education}} \bibinfo{volume}{39}, \bibinfo{number}{8} (\bibinfo{year}{2014}), \bibinfo{pages}{916--931}.
\newblock


\bibitem[Beckett(2002)]%
        {beckett2002teacher}
\bibfield{author}{\bibinfo{person}{Gulbahar Beckett}.} \bibinfo{year}{2002}\natexlab{}.
\newblock \showarticletitle{Teacher and student evaluations of project-based instruction}.
\newblock \bibinfo{journal}{\emph{TESL Canada journal}} (\bibinfo{year}{2002}), \bibinfo{pages}{52--66}.
\newblock


\bibitem[Bender(2012)]%
        {bender2012project}
\bibfield{author}{\bibinfo{person}{William~N Bender}.} \bibinfo{year}{2012}\natexlab{}.
\newblock \bibinfo{booktitle}{\emph{Project-based learning: Differentiating instruction for the 21st century}}.
\newblock \bibinfo{publisher}{Corwin Press}.
\newblock


\bibitem[Bernabei et~al\mbox{.}(2023)]%
        {bernabei2023students}
\bibfield{author}{\bibinfo{person}{Margherita Bernabei}, \bibinfo{person}{Silvia Colabianchi}, \bibinfo{person}{Andrea Falegnami}, {and} \bibinfo{person}{Francesco Costantino}.} \bibinfo{year}{2023}\natexlab{}.
\newblock \showarticletitle{Students’ use of large language models in engineering education: A case study on technology acceptance, perceptions, efficacy, and detection chances}.
\newblock \bibinfo{journal}{\emph{Computers and Education: Artificial Intelligence}}  \bibinfo{volume}{5} (\bibinfo{year}{2023}), \bibinfo{pages}{100172}.
\newblock


\bibitem[Blumenfeld et~al\mbox{.}(1991)]%
        {phyllis1991motivating}
\bibfield{author}{\bibinfo{person}{Phyllis~C. Blumenfeld}, \bibinfo{person}{Elliot Soloway}, \bibinfo{person}{Ronald~W. Marx}, \bibinfo{person}{Joseph~S. Krajcik}, \bibinfo{person}{Mark Guzdial}, {and} \bibinfo{person}{Annemarie Palincsar}.} \bibinfo{year}{1991}\natexlab{}.
\newblock \showarticletitle{{Motivating Project-Based Learning: Sustaining the Doing, Supporting the Learning}}.
\newblock \bibinfo{journal}{\emph{Educational Psychologist}} \bibinfo{volume}{26}, \bibinfo{number}{3-4} (\bibinfo{year}{1991}), \bibinfo{pages}{369--398}.
\newblock


\bibitem[Brassler and Dettmers(2017)]%
        {brassler2017enhance}
\bibfield{author}{\bibinfo{person}{Mirjam Brassler} {and} \bibinfo{person}{Jan Dettmers}.} \bibinfo{year}{2017}\natexlab{}.
\newblock \showarticletitle{{How to Enhance Interdisciplinary Competence—Interdisciplinary Problem-Based Learning versus Interdisciplinary Project-Based Learning}}.
\newblock \bibinfo{journal}{\emph{Interdisciplinary Journal of Problem-Based Learning}} \bibinfo{volume}{11}, \bibinfo{number}{2} (\bibinfo{year}{2017}), \bibinfo{pages}{369--398}.
\newblock


\bibitem[Brodie and Gibbings(2009)]%
        {brodie2009comparison}
\bibfield{author}{\bibinfo{person}{Lyn Brodie} {and} \bibinfo{person}{Peter Gibbings}.} \bibinfo{year}{2009}\natexlab{}.
\newblock \showarticletitle{Comparison of PBL assessment rubrics}. In \bibinfo{booktitle}{\emph{Proceedings of the Research in Engineering Education Symposium (REES 2009)}}.
\newblock


\bibitem[Chen et~al\mbox{.}(2024a)]%
        {chen2024benchmarking}
\bibfield{author}{\bibinfo{person}{Jiawei Chen}, \bibinfo{person}{Hongyu Lin}, \bibinfo{person}{Xianpei Han}, {and} \bibinfo{person}{Le Sun}.} \bibinfo{year}{2024}\natexlab{a}.
\newblock \showarticletitle{Benchmarking large language models in retrieval-augmented generation}. In \bibinfo{booktitle}{\emph{Proceedings of the AAAI Conference on Artificial Intelligence}}, Vol.~\bibinfo{volume}{38}. \bibinfo{pages}{17754--17762}.
\newblock


\bibitem[Chen et~al\mbox{.}(2024b)]%
        {chen2024learning}
\bibfield{author}{\bibinfo{person}{John Chen}, \bibinfo{person}{Xi Lu}, \bibinfo{person}{Yuzhou Du}, \bibinfo{person}{Michael Rejtig}, \bibinfo{person}{Ruth Bagley}, \bibinfo{person}{Mike Horn}, {and} \bibinfo{person}{Uri Wilensky}.} \bibinfo{year}{2024}\natexlab{b}.
\newblock \showarticletitle{Learning agent-based modeling with LLM companions: Experiences of novices and experts using ChatGPT \& NetLogo chat}. In \bibinfo{booktitle}{\emph{Proceedings of the 2024 CHI Conference on Human Factors in Computing Systems}}. \bibinfo{pages}{1--18}.
\newblock


\bibitem[Chen et~al\mbox{.}(2022)]%
        {chen2022effect}
\bibfield{author}{\bibinfo{person}{Shih-Yeh Chen}, \bibinfo{person}{Chin-Feng Lai}, \bibinfo{person}{Ying-Hsun Lai}, {and} \bibinfo{person}{Yu-Sheng Su}.} \bibinfo{year}{2022}\natexlab{}.
\newblock \showarticletitle{Effect of project-based learning on development of students’ creative thinking}.
\newblock \bibinfo{journal}{\emph{The International Journal of Electrical Engineering \& Education}} \bibinfo{volume}{59}, \bibinfo{number}{3} (\bibinfo{year}{2022}), \bibinfo{pages}{232--250}.
\newblock


\bibitem[Chen et~al\mbox{.}(2025)]%
        {chen2024stugptviz}
\bibfield{author}{\bibinfo{person}{Zixin Chen}, \bibinfo{person}{Jiachen Wang}, \bibinfo{person}{Meng Xia}, \bibinfo{person}{Kento Shigyo}, \bibinfo{person}{Dingdong Liu}, \bibinfo{person}{Rong Zhang}, {and} \bibinfo{person}{Huamin Qu}.} \bibinfo{year}{2025}\natexlab{}.
\newblock \showarticletitle{{StuGPTViz: A Visual Analytics Approach to Understand Student-ChatGPT Interactions}}.
\newblock \bibinfo{journal}{\emph{IEEE Transactions on Visualization and Computer Graphics}} \bibinfo{volume}{31}, \bibinfo{number}{1} (\bibinfo{year}{2025}), \bibinfo{pages}{908--918}.
\newblock


\bibitem[Clary et~al\mbox{.}(2011)]%
        {clary2011measuring}
\bibfield{author}{\bibinfo{person}{Renee~M. Clary}, \bibinfo{person}{Robert~F. Brzuszek}, {and} \bibinfo{person}{C.~Taze Fulford}.} \bibinfo{year}{2011}\natexlab{}.
\newblock \showarticletitle{{Measuring Creativity: A Case Study Probing Rubric Effectiveness for Evaluation of Project-Based Learning Solutions}}.
\newblock \bibinfo{journal}{\emph{Creative Education}} \bibinfo{volume}{02}, \bibinfo{number}{04} (\bibinfo{year}{2011}), \bibinfo{pages}{333--340}.
\newblock


\bibitem[Coetzee et~al\mbox{.}(2015)]%
        {coetzee2015structuring}
\bibfield{author}{\bibinfo{person}{Derrick Coetzee}, \bibinfo{person}{Seongtaek Lim}, \bibinfo{person}{Armando Fox}, \bibinfo{person}{Bjorn Hartmann}, {and} \bibinfo{person}{Marti~A Hearst}.} \bibinfo{year}{2015}\natexlab{}.
\newblock \showarticletitle{Structuring interactions for large-scale synchronous peer learning}. In \bibinfo{booktitle}{\emph{Proceedings of the 18th ACM Conference on Computer Supported Cooperative Work \& Social Computing}}. \bibinfo{pages}{1139--1152}.
\newblock


\bibitem[Daryanto et~al\mbox{.}(2025)]%
        {daryanto2025conversate}
\bibfield{author}{\bibinfo{person}{Taufiq Daryanto}, \bibinfo{person}{Xiaohan Ding}, \bibinfo{person}{Lance~T Wilhelm}, \bibinfo{person}{Sophia Stil}, \bibinfo{person}{Kirk~McInnis Knutsen}, {and} \bibinfo{person}{Eugenia~H Rho}.} \bibinfo{year}{2025}\natexlab{}.
\newblock \showarticletitle{Conversate: Supporting Reflective Learning in Interview Practice Through Interactive Simulation and Dialogic Feedback}.
\newblock \bibinfo{journal}{\emph{Proceedings of the ACM on Human-Computer Interaction}} \bibinfo{volume}{9}, \bibinfo{number}{1} (\bibinfo{year}{2025}), \bibinfo{pages}{1--32}.
\newblock


\bibitem[de~Winter et~al\mbox{.}(2023)]%
        {de2023chatgpt}
\bibfield{author}{\bibinfo{person}{Joost C.~F. de Winter}, \bibinfo{person}{Dimitra Dodou}, {and} \bibinfo{person}{Arno H.~A. Stienen}.} \bibinfo{year}{2023}\natexlab{}.
\newblock \showarticletitle{{ChatGPT in Education: Empowering Educators through Methods for Recognition and Assessment}}.
\newblock \bibinfo{journal}{\emph{Informatics}} \bibinfo{volume}{10}, \bibinfo{number}{4}, \bibinfo{pages}{87}.
\newblock


\bibitem[Fan et~al\mbox{.}(2024)]%
        {fan2024lessonplanner}
\bibfield{author}{\bibinfo{person}{Haoxiang Fan}, \bibinfo{person}{Guanzheng Chen}, \bibinfo{person}{Xingbo Wang}, {and} \bibinfo{person}{Zhenhui Peng}.} \bibinfo{year}{2024}\natexlab{}.
\newblock \showarticletitle{LessonPlanner: Assisting novice teachers to prepare pedagogy-driven lesson plans with large language models}. In \bibinfo{booktitle}{\emph{Proceedings of the 37th Annual ACM Symposium on User Interface Software and Technology}}. \bibinfo{pages}{1--20}.
\newblock


\bibitem[Fan et~al\mbox{.}(2025)]%
        {fan2025litlinker}
\bibfield{author}{\bibinfo{person}{Haoxiang Fan}, \bibinfo{person}{Changshuang Zhou}, \bibinfo{person}{Hao Yu}, \bibinfo{person}{Xueyang Wu}, \bibinfo{person}{Jiangyu Gu}, {and} \bibinfo{person}{Zhenhui Peng}.} \bibinfo{year}{2025}\natexlab{}.
\newblock \showarticletitle{LitLinker: Supporting the Ideation of Interdisciplinary Contexts with Large Language Models for Teaching Literature in Elementary Schools}. In \bibinfo{booktitle}{\emph{Proceedings of the 2025 CHI conference on human factors in computing systems}}.
\newblock


\bibitem[Farrelly and Baker(2023)]%
        {farrelly2023generative}
\bibfield{author}{\bibinfo{person}{Tom Farrelly} {and} \bibinfo{person}{Nick Baker}.} \bibinfo{year}{2023}\natexlab{}.
\newblock \showarticletitle{{Generative Artificial Intelligence: Implications and Considerations for Higher Education Practice}}.
\newblock \bibinfo{journal}{\emph{Education Sciences}} \bibinfo{volume}{13}, \bibinfo{number}{11} (\bibinfo{year}{2023}), \bibinfo{pages}{2227--7102}.
\newblock


\bibitem[Fedosejev(2015)]%
        {fedosejev2015react}
\bibfield{author}{\bibinfo{person}{Artemij Fedosejev}.} \bibinfo{year}{2015}\natexlab{}.
\newblock \bibinfo{booktitle}{\emph{React. js essentials}}.
\newblock \bibinfo{publisher}{Packt Publishing Ltd}.
\newblock


\bibitem[Glassman et~al\mbox{.}(2016)]%
        {glassman2016learnersourcing}
\bibfield{author}{\bibinfo{person}{Elena~L Glassman}, \bibinfo{person}{Aaron Lin}, \bibinfo{person}{Carrie~J Cai}, {and} \bibinfo{person}{Robert~C Miller}.} \bibinfo{year}{2016}\natexlab{}.
\newblock \showarticletitle{Learnersourcing personalized hints}. In \bibinfo{booktitle}{\emph{Proceedings of the 19th ACM conference on computer-supported cooperative work \& social computing}}. \bibinfo{pages}{1626--1636}.
\newblock


\bibitem[Grinberg(2018)]%
        {grinberg2018flask}
\bibfield{author}{\bibinfo{person}{Miguel Grinberg}.} \bibinfo{year}{2018}\natexlab{}.
\newblock \bibinfo{booktitle}{\emph{Flask web development: developing web applications with python}}.
\newblock \bibinfo{publisher}{" O'Reilly Media, Inc."}.
\newblock


\bibitem[Guo et~al\mbox{.}(2020)]%
        {guo2020review}
\bibfield{author}{\bibinfo{person}{Pengyue Guo}, \bibinfo{person}{Nadira Saab}, \bibinfo{person}{Lysanne~S Post}, {and} \bibinfo{person}{Wilfried Admiraal}.} \bibinfo{year}{2020}\natexlab{}.
\newblock \showarticletitle{A review of project-based learning in higher education: Student outcomes and measures}.
\newblock \bibinfo{journal}{\emph{International journal of educational research}}  \bibinfo{volume}{102} (\bibinfo{year}{2020}), \bibinfo{pages}{101586}.
\newblock


\bibitem[Hashemi et~al\mbox{.}(2024)]%
        {hashemi2024llm}
\bibfield{author}{\bibinfo{person}{Helia Hashemi}, \bibinfo{person}{Jason Eisner}, \bibinfo{person}{Corby Rosset}, \bibinfo{person}{Benjamin Van~Durme}, {and} \bibinfo{person}{Chris Kedzie}.} \bibinfo{year}{2024}\natexlab{}.
\newblock \showarticletitle{LLM-Rubric: A Multidimensional, Calibrated Approach to Automated Evaluation of Natural Language Texts}. In \bibinfo{booktitle}{\emph{Proceedings of the 62nd Annual Meeting of the Association for Computational Linguistics (Volume 1: Long Papers)}}. \bibinfo{pages}{13806--13834}.
\newblock


\bibitem[Hooper et~al\mbox{.}(2024)]%
        {hooper2024advancing}
\bibfield{author}{\bibinfo{person}{Steffan Hooper}, \bibinfo{person}{Burkhard~C W{\"u}nsche}, \bibinfo{person}{Andrew Luxton-Reilly}, \bibinfo{person}{Paul Denny}, {and} \bibinfo{person}{Tony~Haoran Feng}.} \bibinfo{year}{2024}\natexlab{}.
\newblock \showarticletitle{Advancing Automated Assessment Tools-Opportunities for Innovations in Upper-level Computing Courses: A Position Paper}. In \bibinfo{booktitle}{\emph{Proceedings of the 55th ACM Technical Symposium on Computer Science Education V. 1}}. \bibinfo{pages}{519--525}.
\newblock


\bibitem[Hu et~al\mbox{.}(2024)]%
        {hu2024teaching}
\bibfield{author}{\bibinfo{person}{Bihao Hu}, \bibinfo{person}{Longwei Zheng}, \bibinfo{person}{Jiayi Zhu}, \bibinfo{person}{Lishan Ding}, \bibinfo{person}{Yilei Wang}, {and} \bibinfo{person}{Xiaoqing Gu}.} \bibinfo{year}{2024}\natexlab{}.
\newblock \showarticletitle{Teaching Plan Generation and Evaluation With GPT-4: Unleashing the Potential of LLM in Instructional Design}.
\newblock \bibinfo{journal}{\emph{IEEE Transactions on Learning Technologies}}  \bibinfo{volume}{17} (\bibinfo{year}{2024}), \bibinfo{pages}{1445--1459}.
\newblock


\bibitem[Jia et~al\mbox{.}(2024)]%
        {jia2024llm}
\bibfield{author}{\bibinfo{person}{Qinjin Jia}, \bibinfo{person}{Jialin Cui}, \bibinfo{person}{Haoze Du}, \bibinfo{person}{Parvez Rashid}, \bibinfo{person}{Ruijie Xi}, \bibinfo{person}{Ruochi Li}, {and} \bibinfo{person}{Edward Gehringer}.} \bibinfo{year}{2024}\natexlab{}.
\newblock \showarticletitle{LLM-generated Feedback in Real Classes and Beyond: Perspectives from Students and Instructors}. In \bibinfo{booktitle}{\emph{Proceedings of the 17th International Conference on Educational Data Mining}}. \bibinfo{pages}{862--867}.
\newblock


\bibitem[Jin et~al\mbox{.}(2024)]%
        {jin2024teach}
\bibfield{author}{\bibinfo{person}{Hyoungwook Jin}, \bibinfo{person}{Seonghee Lee}, \bibinfo{person}{Hyungyu Shin}, {and} \bibinfo{person}{Juho Kim}.} \bibinfo{year}{2024}\natexlab{}.
\newblock \showarticletitle{Teach ai how to code: Using large language models as teachable agents for programming education}. In \bibinfo{booktitle}{\emph{Proceedings of the 2024 CHI Conference on Human Factors in Computing Systems}}. \bibinfo{pages}{1--28}.
\newblock


\bibitem[Kazemitabaar et~al\mbox{.}(2024)]%
        {kazemitabaar2024codeaid}
\bibfield{author}{\bibinfo{person}{Majeed Kazemitabaar}, \bibinfo{person}{Runlong Ye}, \bibinfo{person}{Xiaoning Wang}, \bibinfo{person}{Austin~Zachary Henley}, \bibinfo{person}{Paul Denny}, \bibinfo{person}{Michelle Craig}, {and} \bibinfo{person}{Tovi Grossman}.} \bibinfo{year}{2024}\natexlab{}.
\newblock \showarticletitle{Codeaid: Evaluating a classroom deployment of an llm-based programming assistant that balances student and educator needs}. In \bibinfo{booktitle}{\emph{Proceedings of the CHI Conference on Human Factors in Computing Systems}}. \bibinfo{pages}{1--20}.
\newblock


\bibitem[Kokotsaki et~al\mbox{.}(2016)]%
        {dimitra2016project}
\bibfield{author}{\bibinfo{person}{Dimitra Kokotsaki}, \bibinfo{person}{Victoria Menzies}, {and} \bibinfo{person}{Andy Wiggins}.} \bibinfo{year}{2016}\natexlab{}.
\newblock \showarticletitle{Project-based learning: A review of the literature}.
\newblock \bibinfo{journal}{\emph{Improving Schools}} \bibinfo{volume}{19}, \bibinfo{number}{3} (\bibinfo{year}{2016}), \bibinfo{pages}{267--277}.
\newblock
\urldef\tempurl%
\url{https://doi.org/10.1177/1365480216659733}
\showDOI{\tempurl}
\showeprint{https://doi.org/10.1177/1365480216659733}


\bibitem[Kostic et~al\mbox{.}(2024)]%
        {kostic2024llms}
\bibfield{author}{\bibinfo{person}{Milan Kostic}, \bibinfo{person}{Hans~Friedrich Witschel}, \bibinfo{person}{Knut Hinkelmann}, {and} \bibinfo{person}{Maja Spahic-Bogdanovic}.} \bibinfo{year}{2024}\natexlab{}.
\newblock \showarticletitle{LLMs in Automated Essay Evaluation: A Case Study}. In \bibinfo{booktitle}{\emph{Proceedings of the AAAI Symposium Series}}, Vol.~\bibinfo{volume}{3}. \bibinfo{pages}{143--147}.
\newblock


\bibitem[Krajcik and Blumenfeld(2006)]%
        {krajcik2006project}
\bibfield{author}{\bibinfo{person}{Joseph~S Krajcik} {and} \bibinfo{person}{Phyllis~C Blumenfeld}.} \bibinfo{year}{2006}\natexlab{}.
\newblock \bibinfo{booktitle}{\emph{Project-based learning}}.
\newblock \bibinfo{publisher}{na}.
\newblock


\bibitem[Kumar et~al\mbox{.}(2024)]%
        {kumar2024guiding}
\bibfield{author}{\bibinfo{person}{Harsh Kumar}, \bibinfo{person}{Ilya Musabirov}, \bibinfo{person}{Mohi Reza}, \bibinfo{person}{Jiakai Shi}, \bibinfo{person}{Xinyuan Wang}, \bibinfo{person}{Joseph~Jay Williams}, \bibinfo{person}{Anastasia Kuzminykh}, {and} \bibinfo{person}{Michael Liut}.} \bibinfo{year}{2024}\natexlab{}.
\newblock \showarticletitle{Guiding Students in Using LLMs in Supported Learning Environments: Effects on Interaction Dynamics, Learner Performance, Confidence, and Trust}.
\newblock \bibinfo{journal}{\emph{Proceedings of the ACM on Human-Computer Interaction}} \bibinfo{volume}{8}, \bibinfo{number}{CSCW2} (\bibinfo{year}{2024}), \bibinfo{pages}{1--30}.
\newblock


\bibitem[Kumar et~al\mbox{.}(2025)]%
        {kumar2025human}
\bibfield{author}{\bibinfo{person}{Harsh Kumar}, \bibinfo{person}{Jonathan Vincentius}, \bibinfo{person}{Ewan Jordan}, {and} \bibinfo{person}{Ashton Anderson}.} \bibinfo{year}{2025}\natexlab{}.
\newblock \showarticletitle{Human creativity in the age of llms: Randomized experiments on divergent and convergent thinking}. In \bibinfo{booktitle}{\emph{Proceedings of the 2025 CHI Conference on Human Factors in Computing Systems}}. \bibinfo{pages}{1--18}.
\newblock


\bibitem[Leckie and Baird(2011)]%
        {leckie2011rater}
\bibfield{author}{\bibinfo{person}{George Leckie} {and} \bibinfo{person}{Jo-Anne Baird}.} \bibinfo{year}{2011}\natexlab{}.
\newblock \showarticletitle{Rater effects on essay scoring: A multilevel analysis of severity drift, central tendency, and rater experience}.
\newblock \bibinfo{journal}{\emph{Journal of Educational Measurement}} \bibinfo{volume}{48}, \bibinfo{number}{4} (\bibinfo{year}{2011}), \bibinfo{pages}{399--418}.
\newblock


\bibitem[Lee et~al\mbox{.}(2025)]%
        {lee2025impact}
\bibfield{author}{\bibinfo{person}{Hao-Ping Lee}, \bibinfo{person}{Advait Sarkar}, \bibinfo{person}{Lev Tankelevitch}, \bibinfo{person}{Ian Drosos}, \bibinfo{person}{Sean Rintel}, \bibinfo{person}{Richard Banks}, {and} \bibinfo{person}{Nicholas Wilson}.} \bibinfo{year}{2025}\natexlab{}.
\newblock \showarticletitle{The impact of generative AI on critical thinking: Self-reported reductions in cognitive effort and confidence effects from a survey of knowledge workers}. In \bibinfo{booktitle}{\emph{Proceedings of the 2025 CHI conference on human factors in computing systems}}. \bibinfo{pages}{1--22}.
\newblock


\bibitem[Leshed et~al\mbox{.}(2009)]%
        {leshed2009visualizing}
\bibfield{author}{\bibinfo{person}{Gilly Leshed}, \bibinfo{person}{Diego Perez}, \bibinfo{person}{Jeffrey~T Hancock}, \bibinfo{person}{Dan Cosley}, \bibinfo{person}{Jeremy Birnholtz}, \bibinfo{person}{Soyoung Lee}, \bibinfo{person}{Poppy~L McLeod}, {and} \bibinfo{person}{Geri Gay}.} \bibinfo{year}{2009}\natexlab{}.
\newblock \showarticletitle{Visualizing real-time language-based feedback on teamwork behavior in computer-mediated groups}. In \bibinfo{booktitle}{\emph{Proceedings of the SIGCHI Conference on Human Factors in Computing Systems}}. \bibinfo{pages}{537--546}.
\newblock


\bibitem[Li et~al\mbox{.}(2023)]%
        {li2023adapting}
\bibfield{author}{\bibinfo{person}{Qingyao Li}, \bibinfo{person}{Lingyue Fu}, \bibinfo{person}{Weiming Zhang}, \bibinfo{person}{Xianyu Chen}, \bibinfo{person}{Jingwei Yu}, \bibinfo{person}{Wei Xia}, \bibinfo{person}{Weinan Zhang}, \bibinfo{person}{Ruiming Tang}, {and} \bibinfo{person}{Yong Yu}.} \bibinfo{year}{2023}\natexlab{}.
\newblock \showarticletitle{Adapting large language models for education: Foundational capabilities, potentials, and challenges}.
\newblock \bibinfo{journal}{\emph{arXiv preprint arXiv:2401.08664}} (\bibinfo{year}{2023}).
\newblock


\bibitem[Liao et~al\mbox{.}(2023)]%
        {liao2023deepthinkingmap}
\bibfield{author}{\bibinfo{person}{Jingxian Liao}, \bibinfo{person}{Mrinalini Singh}, {and} \bibinfo{person}{Hao-Chuan Wang}.} \bibinfo{year}{2023}\natexlab{}.
\newblock \showarticletitle{Deepthinkingmap: Collaborative video reflection system with graph-based summarizing and commenting}. In \bibinfo{booktitle}{\emph{Companion Publication of the 2023 Conference on Computer Supported Cooperative Work and Social Computing}}. \bibinfo{pages}{369--371}.
\newblock


\bibitem[Liu et~al\mbox{.}(2023)]%
        {liu2023future}
\bibfield{author}{\bibinfo{person}{Ming Liu}, \bibinfo{person}{Yiling Ren}, \bibinfo{person}{Lucy~Michael Nyagoga}, \bibinfo{person}{Francis Stonier}, \bibinfo{person}{Zhongming Wu}, {and} \bibinfo{person}{Liang Yu}.} \bibinfo{year}{2023}\natexlab{}.
\newblock \showarticletitle{Future of education in the era of generative artificial intelligence: Consensus among Chinese scholars on applications of ChatGPT in schools}.
\newblock \bibinfo{journal}{\emph{Future in Educational Research}} \bibinfo{volume}{1}, \bibinfo{number}{1} (\bibinfo{year}{2023}), \bibinfo{pages}{72--101}.
\newblock


\bibitem[Liu et~al\mbox{.}(2024)]%
        {liu2024classmeta}
\bibfield{author}{\bibinfo{person}{Ziyi Liu}, \bibinfo{person}{Zhengzhe Zhu}, \bibinfo{person}{Lijun Zhu}, \bibinfo{person}{Enze Jiang}, \bibinfo{person}{Xiyun Hu}, \bibinfo{person}{Kylie~A Peppler}, {and} \bibinfo{person}{Karthik Ramani}.} \bibinfo{year}{2024}\natexlab{}.
\newblock \showarticletitle{Classmeta: Designing interactive virtual classmate to promote VR classroom participation}. In \bibinfo{booktitle}{\emph{Proceedings of the 2024 CHI Conference on Human Factors in Computing Systems}}. \bibinfo{pages}{1--17}.
\newblock


\bibitem[Maida(2011)]%
        {maida2011project}
\bibfield{author}{\bibinfo{person}{Carl~A Maida}.} \bibinfo{year}{2011}\natexlab{}.
\newblock \showarticletitle{Project-based learning: A critical pedagogy for the twenty-first century}.
\newblock \bibinfo{journal}{\emph{Policy Futures in Education}} \bibinfo{volume}{9}, \bibinfo{number}{6} (\bibinfo{year}{2011}), \bibinfo{pages}{759--768}.
\newblock


\bibitem[Masek and Yamin(2011)]%
        {masek2011effect}
\bibfield{author}{\bibinfo{person}{Alias Masek} {and} \bibinfo{person}{Sulaiman Yamin}.} \bibinfo{year}{2011}\natexlab{}.
\newblock \showarticletitle{The effect of problem based learning on critical thinking ability: a theoretical and empirical review}.
\newblock \bibinfo{journal}{\emph{International Review of Social Sciences and Humanities}} \bibinfo{volume}{2}, \bibinfo{number}{1} (\bibinfo{year}{2011}), \bibinfo{pages}{215--221}.
\newblock


\bibitem[McConvey et~al\mbox{.}(2023)]%
        {mcconvey2023human}
\bibfield{author}{\bibinfo{person}{Kelly McConvey}, \bibinfo{person}{Shion Guha}, {and} \bibinfo{person}{Anastasia Kuzminykh}.} \bibinfo{year}{2023}\natexlab{}.
\newblock \showarticletitle{A human-centered review of algorithms in decision-making in higher education}. In \bibinfo{booktitle}{\emph{Proceedings of the 2023 CHI Conference on Human Factors in Computing Systems}}. \bibinfo{pages}{1--15}.
\newblock


\bibitem[Menzies et~al\mbox{.}(2016)]%
        {menzies2016project}
\bibfield{author}{\bibinfo{person}{Victoria Menzies}, \bibinfo{person}{Catherine Hewitt}, \bibinfo{person}{Dimitra Kokotsaki}, \bibinfo{person}{Clare Collyer}, {and} \bibinfo{person}{Andy Wiggins}.} \bibinfo{year}{2016}\natexlab{}.
\newblock \showarticletitle{Project Based Learning: evaluation report and executive summary}.
\newblock  (\bibinfo{year}{2016}).
\newblock


\bibitem[Meyer et~al\mbox{.}(2024)]%
        {meyer2024using}
\bibfield{author}{\bibinfo{person}{Jennifer Meyer}, \bibinfo{person}{Thorben Jansen}, \bibinfo{person}{Ronja Schiller}, \bibinfo{person}{Lucas~W Liebenow}, \bibinfo{person}{Marlene Steinbach}, \bibinfo{person}{Andrea Horbach}, {and} \bibinfo{person}{Johanna Fleckenstein}.} \bibinfo{year}{2024}\natexlab{}.
\newblock \showarticletitle{Using LLMs to bring evidence-based feedback into the classroom: AI-generated feedback increases secondary students’ text revision, motivation, and positive emotions}.
\newblock \bibinfo{journal}{\emph{Computers and Education: Artificial Intelligence}}  \bibinfo{volume}{6} (\bibinfo{year}{2024}), \bibinfo{pages}{100199}.
\newblock


\bibitem[Meyer et~al\mbox{.}(2023)]%
        {meyer2023chatgpt}
\bibfield{author}{\bibinfo{person}{Jesse~G Meyer}, \bibinfo{person}{Ryan~J Urbanowicz}, \bibinfo{person}{Patrick~CN Martin}, \bibinfo{person}{Karen O’Connor}, \bibinfo{person}{Ruowang Li}, \bibinfo{person}{Pei-Chen Peng}, \bibinfo{person}{Tiffani~J Bright}, \bibinfo{person}{Nicholas Tatonetti}, \bibinfo{person}{Kyoung~Jae Won}, \bibinfo{person}{Graciela Gonzalez-Hernandez}, {et~al\mbox{.}}} \bibinfo{year}{2023}\natexlab{}.
\newblock \showarticletitle{ChatGPT and large language models in academia: opportunities and challenges}.
\newblock \bibinfo{journal}{\emph{BioData Mining}} \bibinfo{volume}{16}, \bibinfo{number}{1} (\bibinfo{year}{2023}), \bibinfo{pages}{20}.
\newblock


\bibitem[Moore et~al\mbox{.}(2023)]%
        {moore2023empowering}
\bibfield{author}{\bibinfo{person}{Steven Moore}, \bibinfo{person}{Richard Tong}, \bibinfo{person}{Anjali Singh}, \bibinfo{person}{Zitao Liu}, \bibinfo{person}{Xiangen Hu}, \bibinfo{person}{Yu Lu}, \bibinfo{person}{Joleen Liang}, \bibinfo{person}{Chen Cao}, \bibinfo{person}{Hassan Khosravi}, \bibinfo{person}{Paul Denny}, {et~al\mbox{.}}} \bibinfo{year}{2023}\natexlab{}.
\newblock \showarticletitle{Empowering education with llms-the next-gen interface and content generation}. In \bibinfo{booktitle}{\emph{International Conference on Artificial Intelligence in Education}}. Springer, \bibinfo{pages}{32--37}.
\newblock


\bibitem[Ngoon et~al\mbox{.}(2018)]%
        {ngoon2018interactive}
\bibfield{author}{\bibinfo{person}{Tricia~J Ngoon}, \bibinfo{person}{C~Ailie Fraser}, \bibinfo{person}{Ariel~S Weingarten}, \bibinfo{person}{Mira Dontcheva}, {and} \bibinfo{person}{Scott Klemmer}.} \bibinfo{year}{2018}\natexlab{}.
\newblock \showarticletitle{Interactive guidance techniques for improving creative feedback}. In \bibinfo{booktitle}{\emph{Proceedings of the 2018 CHI Conference on Human Factors in Computing Systems}}. \bibinfo{pages}{1--11}.
\newblock


\bibitem[Nilson and Stanny(2015)]%
        {nilson2015specifications}
\bibfield{author}{\bibinfo{person}{Linda~B Nilson} {and} \bibinfo{person}{Claudia~J Stanny}.} \bibinfo{year}{2015}\natexlab{}.
\newblock \bibinfo{booktitle}{\emph{Specifications grading: Restoring rigor, motivating students, and saving faculty time}}.
\newblock \bibinfo{publisher}{Routledge}.
\newblock


\bibitem[Olney(2023)]%
        {olney2023generating}
\bibfield{author}{\bibinfo{person}{Andrew~M Olney}.} \bibinfo{year}{2023}\natexlab{}.
\newblock \showarticletitle{Generating Multiple Choice Questions from a Textbook: LLMs Match Human Performance on Most Metrics.}. In \bibinfo{booktitle}{\emph{LLM@ AIED}}. \bibinfo{pages}{111--128}.
\newblock


\bibitem[Oney et~al\mbox{.}(2018)]%
        {oney2018creating}
\bibfield{author}{\bibinfo{person}{Steve Oney}, \bibinfo{person}{Christopher Brooks}, {and} \bibinfo{person}{Paul Resnick}.} \bibinfo{year}{2018}\natexlab{}.
\newblock \showarticletitle{Creating guided code explanations with chat. codes}.
\newblock \bibinfo{journal}{\emph{Proceedings of the ACM on Human-Computer Interaction}} \bibinfo{volume}{2}, \bibinfo{number}{CSCW} (\bibinfo{year}{2018}), \bibinfo{pages}{1--20}.
\newblock


\bibitem[OpenAI(2024a)]%
        {filesearch2024gpt}
\bibfield{author}{\bibinfo{person}{OpenAI}.} \bibinfo{year}{2024}\natexlab{a}.
\newblock \bibinfo{title}{Assistants API: File Search}.
\newblock \bibinfo{howpublished}{\url{https://platform.openai.com/docs/assistants/tools/file-search}}.
\newblock
\newblock
\shownote{[Online]}.


\bibitem[OpenAI(2024b)]%
        {assistant2024gpt}
\bibfield{author}{\bibinfo{person}{OpenAI}.} \bibinfo{year}{2024}\natexlab{b}.
\newblock \bibinfo{title}{Assistants API Overview}.
\newblock \bibinfo{howpublished}{\url{https://platform.openai.com/docs/assistants/overview}}.
\newblock
\newblock
\shownote{[Online]}.


\bibitem[OpenAI(2024c)]%
        {structured2024gpt}
\bibfield{author}{\bibinfo{person}{OpenAI}.} \bibinfo{year}{2024}\natexlab{c}.
\newblock \bibinfo{title}{CAPABILITIES: Structured Outputs}.
\newblock \bibinfo{howpublished}{\url{https://platform.openai.com/docs/guides/structured-outputs}}.
\newblock
\newblock
\shownote{[Online]}.


\bibitem[OpenAI(2024d)]%
        {gpt2024}
\bibfield{author}{\bibinfo{person}{OpenAI}.} \bibinfo{year}{2024}\natexlab{d}.
\newblock \bibinfo{title}{GPT-4o System Card}.
\newblock \bibinfo{howpublished}{\url{https://openai.com/index/gpt-4o-system-card/}, journal={GPT-4O system card | openai}}.
\newblock
\newblock
\shownote{[Online; Accessed 08-August-2024]}.


\bibitem[Pan et~al\mbox{.}(2025)]%
        {pan2025tutorup}
\bibfield{author}{\bibinfo{person}{Sitong Pan}, \bibinfo{person}{Robin Schmucker}, \bibinfo{person}{Bernardo Garcia~Bulle Bueno}, \bibinfo{person}{Salome~Aguilar Llanes}, \bibinfo{person}{Fernanda~Albo Alarc{\'o}n}, \bibinfo{person}{Hangxiao Zhu}, \bibinfo{person}{Adam Teo}, {and} \bibinfo{person}{Meng Xia}.} \bibinfo{year}{2025}\natexlab{}.
\newblock \showarticletitle{TutorUp: What If Your Students Were Simulated? Training Tutors to Address Engagement Challenges in Online Learning}. In \bibinfo{booktitle}{\emph{Proceedings of the 2025 CHI conference on human factors in computing systems}}.
\newblock


\bibitem[Panadero and Jonsson(2020)]%
        {panadero2020critical}
\bibfield{author}{\bibinfo{person}{Ernesto Panadero} {and} \bibinfo{person}{Anders Jonsson}.} \bibinfo{year}{2020}\natexlab{}.
\newblock \showarticletitle{A critical review of the arguments against the use of rubrics}.
\newblock \bibinfo{journal}{\emph{Educational Research Review}}  \bibinfo{volume}{30} (\bibinfo{year}{2020}), \bibinfo{pages}{100329}.
\newblock


\bibitem[Panickssery et~al\mbox{.}(2024)]%
        {panickssery2024llm}
\bibfield{author}{\bibinfo{person}{Arjun Panickssery}, \bibinfo{person}{Samuel Bowman}, {and} \bibinfo{person}{Shi Feng}.} \bibinfo{year}{2024}\natexlab{}.
\newblock \showarticletitle{Llm evaluators recognize and favor their own generations}.
\newblock \bibinfo{journal}{\emph{Advances in Neural Information Processing Systems}}  \bibinfo{volume}{37} (\bibinfo{year}{2024}), \bibinfo{pages}{68772--68802}.
\newblock


\bibitem[Park et~al\mbox{.}(2017)]%
        {park2017eliph}
\bibfield{author}{\bibinfo{person}{Jungkook Park}, \bibinfo{person}{Yeong~Hoon Park}, \bibinfo{person}{Suin Kim}, {and} \bibinfo{person}{Alice Oh}.} \bibinfo{year}{2017}\natexlab{}.
\newblock \showarticletitle{Eliph: Effective visualization of code history for peer assessment in programming education}. In \bibinfo{booktitle}{\emph{Proceedings of the 2017 ACM Conference on Computer Supported Cooperative Work and Social Computing}}. \bibinfo{pages}{458--467}.
\newblock


\bibitem[Rasooli et~al\mbox{.}(2018)]%
        {rasooli2018re}
\bibfield{author}{\bibinfo{person}{Amirhossein Rasooli}, \bibinfo{person}{Hamed Zandi}, {and} \bibinfo{person}{Christopher DeLuca}.} \bibinfo{year}{2018}\natexlab{}.
\newblock \showarticletitle{Re-conceptualizing classroom assessment fairness: A systematic meta-ethnography of assessment literature and beyond}.
\newblock \bibinfo{journal}{\emph{Studies in Educational Evaluation}}  \bibinfo{volume}{56} (\bibinfo{year}{2018}), \bibinfo{pages}{164--181}.
\newblock


\bibitem[Ravi et~al\mbox{.}(2025)]%
        {ravi2025co}
\bibfield{author}{\bibinfo{person}{Prerna Ravi}, \bibinfo{person}{John Masla}, \bibinfo{person}{Gisella Kakoti}, \bibinfo{person}{Grace Lin}, \bibinfo{person}{Emma Anderson}, \bibinfo{person}{Matt Taylor}, \bibinfo{person}{Anastasia Ostrowski}, \bibinfo{person}{Cynthia Breazeal}, \bibinfo{person}{Eric Klopfer}, {and} \bibinfo{person}{Hal Abelson}.} \bibinfo{year}{2025}\natexlab{}.
\newblock \showarticletitle{Co-designing Large Language Model Tools for Project-Based Learning with K12 Educators}. In \bibinfo{booktitle}{\emph{Proceedings of the 2025 CHI conference on human factors in computing systems}}.
\newblock


\bibitem[Ross et~al\mbox{.}(2023)]%
        {ross2023programmer}
\bibfield{author}{\bibinfo{person}{Steven~I Ross}, \bibinfo{person}{Fernando Martinez}, \bibinfo{person}{Stephanie Houde}, \bibinfo{person}{Michael Muller}, {and} \bibinfo{person}{Justin~D Weisz}.} \bibinfo{year}{2023}\natexlab{}.
\newblock \showarticletitle{The programmer’s assistant: Conversational interaction with a large language model for software development}. In \bibinfo{booktitle}{\emph{Proceedings of the 28th International Conference on Intelligent User Interfaces}}. \bibinfo{pages}{491--514}.
\newblock


\bibitem[Saliba et~al\mbox{.}(2024)]%
        {saliba2024learning}
\bibfield{author}{\bibinfo{person}{Liam Saliba}, \bibinfo{person}{Elisa Shioji}, \bibinfo{person}{Eduardo Oliveira}, \bibinfo{person}{Shaanan Cohney}, {and} \bibinfo{person}{Jianzhong Qi}.} \bibinfo{year}{2024}\natexlab{}.
\newblock \showarticletitle{Learning with Style: Improving Student Code-Style Through Better Automated Feedback}. In \bibinfo{booktitle}{\emph{Proceedings of the 55th ACM Technical Symposium on Computer Science Education V. 1}}. \bibinfo{pages}{1175--1181}.
\newblock


\bibitem[Sasson et~al\mbox{.}(2018)]%
        {sasson2018fostering}
\bibfield{author}{\bibinfo{person}{Irit Sasson}, \bibinfo{person}{Itamar Yehuda}, {and} \bibinfo{person}{Noam Malkinson}.} \bibinfo{year}{2018}\natexlab{}.
\newblock \showarticletitle{Fostering the skills of critical thinking and question-posing in a project-based learning environment}.
\newblock \bibinfo{journal}{\emph{Thinking Skills and Creativity}}  \bibinfo{volume}{29} (\bibinfo{year}{2018}), \bibinfo{pages}{203--212}.
\newblock


\bibitem[Schank and Hamel(2004)]%
        {schank2004collaborative}
\bibfield{author}{\bibinfo{person}{Patricia Schank} {and} \bibinfo{person}{Lawrence Hamel}.} \bibinfo{year}{2004}\natexlab{}.
\newblock \showarticletitle{Collaborative modeling: hiding UML and promoting data examples in NEMo}. In \bibinfo{booktitle}{\emph{Proceedings of the 2004 ACM conference on Computer supported cooperative work}}. \bibinfo{pages}{574--577}.
\newblock


\bibitem[Sedgwick(2012)]%
        {sedgwick2012pearson}
\bibfield{author}{\bibinfo{person}{Philip Sedgwick}.} \bibinfo{year}{2012}\natexlab{}.
\newblock \showarticletitle{Pearson’s correlation coefficient}.
\newblock \bibinfo{journal}{\emph{Bmj}}  \bibinfo{volume}{345} (\bibinfo{year}{2012}).
\newblock


\bibitem[Sedgwick(2014)]%
        {sedgwick2014spearman}
\bibfield{author}{\bibinfo{person}{Philip Sedgwick}.} \bibinfo{year}{2014}\natexlab{}.
\newblock \showarticletitle{Spearman’s rank correlation coefficient}.
\newblock \bibinfo{journal}{\emph{Bmj}}  \bibinfo{volume}{349} (\bibinfo{year}{2014}).
\newblock


\bibitem[Shao et~al\mbox{.}(2025)]%
        {shao2025unlocking}
\bibfield{author}{\bibinfo{person}{Zekai Shao}, \bibinfo{person}{Siyu Yuan}, \bibinfo{person}{Lin Gao}, \bibinfo{person}{Yixuan He}, \bibinfo{person}{Deqing Yang}, {and} \bibinfo{person}{Siming Chen}.} \bibinfo{year}{2025}\natexlab{}.
\newblock \showarticletitle{Unlocking Scientific Concepts: How Effective Are LLM-Generated Analogies for Student Understanding and Classroom Practice?}. In \bibinfo{booktitle}{\emph{Proceedings of the 2025 CHI conference on human factors in computing systems}}.
\newblock


\bibitem[Sidorkin(2024)]%
        {sidorkin2024embracing}
\bibfield{author}{\bibinfo{person}{Alexander~M Sidorkin}.} \bibinfo{year}{2024}\natexlab{}.
\newblock \bibinfo{booktitle}{\emph{Embracing chatbots in higher education: the use of artificial intelligence in teaching, administration, and scholarship}}.
\newblock \bibinfo{publisher}{Taylor \& Francis}.
\newblock


\bibitem[Song et~al\mbox{.}(2024)]%
        {song2024automated}
\bibfield{author}{\bibinfo{person}{Yishen Song}, \bibinfo{person}{Qianta Zhu}, \bibinfo{person}{Huaibo Wang}, {and} \bibinfo{person}{Qinhua Zheng}.} \bibinfo{year}{2024}\natexlab{}.
\newblock \showarticletitle{Automated Essay Scoring and Revising Based on Open-Source Large Language Models}.
\newblock \bibinfo{journal}{\emph{IEEE Transactions on Learning Technologies}} (\bibinfo{year}{2024}).
\newblock


\bibitem[Stamper et~al\mbox{.}(2024)]%
        {stamper2024enhancing}
\bibfield{author}{\bibinfo{person}{John Stamper}, \bibinfo{person}{Ruiwei Xiao}, {and} \bibinfo{person}{Xinying Hou}.} \bibinfo{year}{2024}\natexlab{}.
\newblock \showarticletitle{Enhancing llm-based feedback: Insights from intelligent tutoring systems and the learning sciences}. In \bibinfo{booktitle}{\emph{International Conference on Artificial Intelligence in Education}}. Springer, \bibinfo{pages}{32--43}.
\newblock


\bibitem[Tang et~al\mbox{.}(2024)]%
        {tang2024vizgroup}
\bibfield{author}{\bibinfo{person}{Xiaohang Tang}, \bibinfo{person}{Sam Wong}, \bibinfo{person}{Kevin Pu}, \bibinfo{person}{Xi Chen}, \bibinfo{person}{Yalong Yang}, {and} \bibinfo{person}{Yan Chen}.} \bibinfo{year}{2024}\natexlab{}.
\newblock \showarticletitle{VizGroup: An AI-assisted Event-driven System for Collaborative Programming Learning Analytics}. In \bibinfo{booktitle}{\emph{Proceedings of the 37th Annual ACM Symposium on User Interface Software and Technology}}. \bibinfo{pages}{1--22}.
\newblock


\bibitem[Tian et~al\mbox{.}(2024)]%
        {tian2024opportunities}
\bibfield{author}{\bibinfo{person}{Shubo Tian}, \bibinfo{person}{Qiao Jin}, \bibinfo{person}{Lana Yeganova}, \bibinfo{person}{Po-Ting Lai}, \bibinfo{person}{Qingqing Zhu}, \bibinfo{person}{Xiuying Chen}, \bibinfo{person}{Yifan Yang}, \bibinfo{person}{Qingyu Chen}, \bibinfo{person}{Won Kim}, \bibinfo{person}{Donald~C Comeau}, {et~al\mbox{.}}} \bibinfo{year}{2024}\natexlab{}.
\newblock \showarticletitle{Opportunities and challenges for ChatGPT and large language models in biomedicine and health}.
\newblock \bibinfo{journal}{\emph{Briefings in Bioinformatics}} \bibinfo{volume}{25}, \bibinfo{number}{1} (\bibinfo{year}{2024}), \bibinfo{pages}{bbad493}.
\newblock


\bibitem[Timmis et~al\mbox{.}(2016)]%
        {timmis2016rethinking}
\bibfield{author}{\bibinfo{person}{Sue Timmis}, \bibinfo{person}{Patricia Broadfoot}, \bibinfo{person}{Rosamund Sutherland}, {and} \bibinfo{person}{Alison Oldfield}.} \bibinfo{year}{2016}\natexlab{}.
\newblock \showarticletitle{Rethinking assessment in a digital age: Opportunities, challenges and risks}.
\newblock \bibinfo{journal}{\emph{British Educational Research Journal}} \bibinfo{volume}{42}, \bibinfo{number}{3} (\bibinfo{year}{2016}), \bibinfo{pages}{454--476}.
\newblock


\bibitem[Tran et~al\mbox{.}(2023)]%
        {tran2023generating}
\bibfield{author}{\bibinfo{person}{Andrew Tran}, \bibinfo{person}{Kenneth Angelikas}, \bibinfo{person}{Egi Rama}, \bibinfo{person}{Chiku Okechukwu}, \bibinfo{person}{David~H Smith}, {and} \bibinfo{person}{Stephen MacNeil}.} \bibinfo{year}{2023}\natexlab{}.
\newblock \showarticletitle{Generating multiple choice questions for computing courses using large language models}. In \bibinfo{booktitle}{\emph{2023 IEEE Frontiers in Education Conference (FIE)}}. IEEE, \bibinfo{pages}{1--8}.
\newblock


\bibitem[Tsai et~al\mbox{.}(2023)]%
        {tsai2023exploring}
\bibfield{author}{\bibinfo{person}{Meng-Lin Tsai}, \bibinfo{person}{Chong~Wei Ong}, {and} \bibinfo{person}{Cheng-Liang Chen}.} \bibinfo{year}{2023}\natexlab{}.
\newblock \showarticletitle{Exploring the use of large language models (LLMs) in chemical engineering education: Building core course problem models with Chat-GPT}.
\newblock \bibinfo{journal}{\emph{Education for Chemical Engineers}}  \bibinfo{volume}{44} (\bibinfo{year}{2023}), \bibinfo{pages}{71--95}.
\newblock


\bibitem[Walvoord and Anderson(2011)]%
        {walvoord2011effective}
\bibfield{author}{\bibinfo{person}{Barbara~E Walvoord} {and} \bibinfo{person}{Virginia~Johnson Anderson}.} \bibinfo{year}{2011}\natexlab{}.
\newblock \bibinfo{booktitle}{\emph{Effective grading: A tool for learning and assessment in college}}.
\newblock \bibinfo{publisher}{John Wiley \& Sons}.
\newblock


\bibitem[Wang et~al\mbox{.}(2020)]%
        {wang2020voicecoach}
\bibfield{author}{\bibinfo{person}{Xingbo Wang}, \bibinfo{person}{Haipeng Zeng}, \bibinfo{person}{Yong Wang}, \bibinfo{person}{Aoyu Wu}, \bibinfo{person}{Zhida Sun}, \bibinfo{person}{Xiaojuan Ma}, {and} \bibinfo{person}{Huamin Qu}.} \bibinfo{year}{2020}\natexlab{}.
\newblock \showarticletitle{Voicecoach: Interactive evidence-based training for voice modulation skills in public speaking}. In \bibinfo{booktitle}{\emph{Proceedings of the 2020 CHI Conference on Human Factors in Computing Systems}}. \bibinfo{pages}{1--12}.
\newblock


\bibitem[Warner et~al\mbox{.}(2023)]%
        {warner2023slidespecs}
\bibfield{author}{\bibinfo{person}{Jeremy Warner}, \bibinfo{person}{Amy Pavel}, \bibinfo{person}{Tonya Nguyen}, \bibinfo{person}{Maneesh Agrawala}, {and} \bibinfo{person}{Bjoern Hartmann}.} \bibinfo{year}{2023}\natexlab{}.
\newblock \showarticletitle{Slidespecs: Automatic and interactive presentation feedback collation}. In \bibinfo{booktitle}{\emph{Proceedings of the 28th International Conference on Intelligent User Interfaces}}. \bibinfo{pages}{695--709}.
\newblock


\bibitem[Wormeli(2023)]%
        {wormeli2023fair}
\bibfield{author}{\bibinfo{person}{Rick Wormeli}.} \bibinfo{year}{2023}\natexlab{}.
\newblock \bibinfo{booktitle}{\emph{Fair isn't always equal: Assessment \& Grading in the Differentiated Classroom}}.
\newblock \bibinfo{publisher}{Routledge}.
\newblock


\bibitem[Yen et~al\mbox{.}(2024)]%
        {yen2024give}
\bibfield{author}{\bibinfo{person}{Yu-Chun~Grace Yen}, \bibinfo{person}{Isabelle~Yan Pan}, \bibinfo{person}{Grace Lin}, \bibinfo{person}{Mingyi Li}, \bibinfo{person}{Hyoungwook Jin}, \bibinfo{person}{Mengyi Chen}, \bibinfo{person}{Haijun Xia}, \bibinfo{person}{Steven~P Dow}, {et~al\mbox{.}}} \bibinfo{year}{2024}\natexlab{}.
\newblock \showarticletitle{When to Give Feedback: Exploring Tradeoffs in the Timing of Design Feedback}.
\newblock  (\bibinfo{year}{2024}).
\newblock


\bibitem[Yoon et~al\mbox{.}(2016)]%
        {yoon2016richreview++}
\bibfield{author}{\bibinfo{person}{Dongwook Yoon}, \bibinfo{person}{Nicholas Chen}, \bibinfo{person}{Bernie Randles}, \bibinfo{person}{Amy Cheatle}, \bibinfo{person}{Corinna~E L{\"o}ckenhoff}, \bibinfo{person}{Steven~J Jackson}, \bibinfo{person}{Abigail Sellen}, {and} \bibinfo{person}{Fran{\c{c}}ois Guimbreti{\`e}re}.} \bibinfo{year}{2016}\natexlab{}.
\newblock \showarticletitle{Richreview++ deployment of a collaborative multi-modal annotation system for instructor feedback and peer discussion}. In \bibinfo{booktitle}{\emph{Proceedings of the 19th ACM Conference on Computer-Supported Cooperative Work \& Social Computing}}. \bibinfo{pages}{195--205}.
\newblock


\bibitem[Yoon and Mitros(2015)]%
        {yoon2015multi}
\bibfield{author}{\bibinfo{person}{Dongwook Yoon} {and} \bibinfo{person}{Piotr Mitros}.} \bibinfo{year}{2015}\natexlab{}.
\newblock \showarticletitle{Multi-modal peer discussion with RichReview on edX}. In \bibinfo{booktitle}{\emph{Adjunct Proceedings of the 28th Annual ACM Symposium on User Interface Software \& Technology}}. \bibinfo{pages}{55--56}.
\newblock


\bibitem[Yuan et~al\mbox{.}(2016)]%
        {yuan2016almost}
\bibfield{author}{\bibinfo{person}{Alvin Yuan}, \bibinfo{person}{Kurt Luther}, \bibinfo{person}{Markus Krause}, \bibinfo{person}{Sophie~Isabel Vennix}, \bibinfo{person}{Steven~P Dow}, {and} \bibinfo{person}{Bjorn Hartmann}.} \bibinfo{year}{2016}\natexlab{}.
\newblock \showarticletitle{Almost an expert: The effects of rubrics and expertise on perceived value of crowdsourced design critiques}. In \bibinfo{booktitle}{\emph{Proceedings of the 19th ACM Conference on Computer-Supported Cooperative Work \& Social Computing}}. \bibinfo{pages}{1005--1017}.
\newblock


\bibitem[Zhang et~al\mbox{.}(2025)]%
        {zhang2025cpvis}
\bibfield{author}{\bibinfo{person}{Gefei Zhang}, \bibinfo{person}{Shenming Ji}, \bibinfo{person}{Yicao Li}, \bibinfo{person}{Jingwei Tang}, \bibinfo{person}{Jihong Ding}, \bibinfo{person}{Meng Xia}, \bibinfo{person}{Guodao Sun}, {and} \bibinfo{person}{Ronghua Liang}.} \bibinfo{year}{2025}\natexlab{}.
\newblock \showarticletitle{CPVis: Evidence-based Multimodal Learning Analytics for Evaluation in Collaborative Programming}. In \bibinfo{booktitle}{\emph{Proceedings of the 2025 CHI conference on human factors in computing systems}}.
\newblock


\bibitem[Zhang et~al\mbox{.}(2024)]%
        {zhang2024peer}
\bibfield{author}{\bibinfo{person}{Liang Zhang}, \bibinfo{person}{Tianyi Chen}, \bibinfo{person}{Yue Zong}, {and} \bibinfo{person}{Xiaopeng Gao}.} \bibinfo{year}{2024}\natexlab{}.
\newblock \showarticletitle{A Peer Grading Approach for Open-ended Programming Projects Based on Binary System and Swiss System}. In \bibinfo{booktitle}{\emph{Proceedings of the 55th ACM Technical Symposium on Computer Science Education V. 1}}. \bibinfo{pages}{1484--1490}.
\newblock


\bibitem[Zheng et~al\mbox{.}(2024)]%
        {zheng2024selfgauge}
\bibfield{author}{\bibinfo{person}{Chengbo Zheng}, \bibinfo{person}{Zeyu Huang}, \bibinfo{person}{Shuai Ma}, {and} \bibinfo{person}{Xiaojuan Ma}.} \bibinfo{year}{2024}\natexlab{}.
\newblock \showarticletitle{SelfGauge: An Intelligent Tool to Support Student Self-assessment in GenAI-enhanced Project-based Learning}. In \bibinfo{booktitle}{\emph{Adjunct Proceedings of the 37th Annual ACM Symposium on User Interface Software and Technology}}. \bibinfo{pages}{1--3}.
\newblock


\bibitem[Zheng et~al\mbox{.}(2023)]%
        {zheng2023competent}
\bibfield{author}{\bibinfo{person}{Chengbo Zheng}, \bibinfo{person}{Yuheng Wu}, \bibinfo{person}{Chuhan Shi}, \bibinfo{person}{Shuai Ma}, \bibinfo{person}{Jiehui Luo}, {and} \bibinfo{person}{Xiaojuan Ma}.} \bibinfo{year}{2023}\natexlab{}.
\newblock \showarticletitle{Competent but Rigid: Identifying the Gap in Empowering AI to Participate Equally in Group Decision-Making}. In \bibinfo{booktitle}{\emph{Proceedings of the 2023 CHI Conference on Human Factors in Computing Systems}}. \bibinfo{pages}{1--19}.
\newblock


\end{thebibliography}
